**Quarter and Full Car Models Optimisation of Passive and Active Suspension System Using Genetic Algorithm**

Srikanth Raju Mudduluru, Mahmoud Chizari

*School of Physics, Engineering and Computer Science, University of Hertfordshire, Hatfield, UK*

**Abstract**

This study evaluates a suspension design of a passenger car to obtain maximum rider's comfort when the vehicle is subjected to different road profile or road surface condition. The challenge will be on finding a balance between the rider's comfort and vehicle handling to optimize design parameters. The study uses a simple passive suspension system and an active suspension model integrated with a pneumatic actuator controlled by proportional integral derivative (PID) controller in both quarter car and full car models having a different degree of freedoms (DOF) and increasing degrees of complexities. The quarter car considered as a 2-DOF model, while the full car model is a 7-DOF model. The design process set to optimise the spring stiffnesses, damping coefficients and actuator PID controller gains. For optimisation, the research featured genetic algorithm optimisation technique to obtain a balanced response of the vehicle as evaluated from the displacement, velocity and acceleration of sprung and unsprung masses along with different human comfort and vehicle performance criteria. The results revealed that the active suspension system with optimised spring stiffness, damping coefficients and PID gains demonstrated the superior riding comfort and road holding compared to a passive suspension system.

**Keywords:** Passive and active suspensions, vehicle comfort, vibration, road profile, optimisation

**Nomenclature**

$Z_s$ -Sprung Mass Displacement

$\dot{Z}_s$-Sprung Mass Velocity

$Z_u$ -Unsprung Mass Displacement

$\dot{Z}_u$ -Unsprung Mass Velocity

$Z_r$ -Road Profile

$Z_{ufl}$ -Front-Left Unsprung Mass Displacement

$\dot{Z}_{ufl}$-Front-Left Unsprung Mass Velocity

$Z_{fr}$ -Front-Right Unsprung Mass Displacement

$\dot{Z}_{ufr}$-Front-Right Unsprung Mass Velocity

$Z_{url}$ -Rear-Left Unsprung Mass Displacement

$\dot{Z}_{url}$-Rear-Left Unsprung Mass Velocity

$Z_{urr}$ -Rear-Right Unsprung Mass Displacement

$\dot{Z}_{urr}$-Rear-Right Unsprung Mass Velocity

$Z_{rfl}$ -Road-Profile at Front-Left Wheel

$Z_{rfr}$ -Road-Profile at Front-Right Wheel

$Z_{rrl}$ -Road-Profile at Rear-Left Wheel

$Z_{rrr}$ -Road-Profile at Right-Right Wheel

$\theta$ -Pitch Angle

$\dot{\theta}$ -Pitch Angular Velocity

$\phi$ -Roll Angle

$\dot{\phi}$ -Roll Angular Velocity

## 1. Introduction

Since the advent of transport technologies, for easy mobility of and comfort humans, vehicle suspension systems (VSS) have played significant roles, from the horse-drawn carriages with flexible leaf springs at the four corners to the modern springs (generally made of steel and its alloys, chrome silicon, chrome vanadium, beryllium copper, phosphor bronze and titanium) in modern automobiles with complex control algorithms [1]. On this account, the significance of VSS is invariably factored into vehicles' design because, in situations where the drivers and passengers have to be exposed to prolonged driving distances on uneven road topographies, they are prone to discomfort and health-related problems [1-3]. In general, vehicle VSS are classified into three sets, namely the passive suspension system (PSS), the semi-active suspension system (SASS), and the active suspension system (ASS) [2]. As far as this work is concerned, only the PSS and ASS will be given considerable emphasis herein. Nonetheless, one important thing to note about SASS is that they have aroused the interest of researchers and received a substantial amount of attention because they apparently provide the best trade-off between economic viability and field-proven reliability when compared with PSS and ASS [3].

On the one hand, a PSS uses damping elements such as hydraulic shock absorber/viscous dampers positioned between the body and wheels of the vehicle to restrain the motion of the body and wheel by limiting their relative velocities to a rate that gives a satisfactory comfort [3]. The PSS only offers a trade-off between ride comfort and road holding by providing spring and damping coefficients with fixed rates [4]. On the other hand, an ASS is a closed-loop system with a feedback signal representing all or a few of the system variables to control actuators. The actuators are added in addition to the usual passive elements, or the passive elements are replaced altogether. In general, the ASS contains external power sources, force-generating actuators, measuring and sensing instruments along with the conditioning and amplifying devices [5]. Nonetheless, it is important to note that irrespective of the suspension type, their primary function is to reduce or eliminate the road excitations (road shocks and vibrations) transmitted to the body of the vehicle, that is, dampen road profile-induced and vehicle manoeuvre-induced vibrational energies from the vehicle which most often than not is invariably discomforting and detrimental to the health and safety of the driver, the passengers and the vehicle itself [1-5]. Owing to these phenomena, a good VSS should provide a comfortable ride and decent road holding within a reasonable array of suspension deflection [3-5].

For a long time now, the design of perfect VSS has remained one of the most daunting challenges for automotive engineers. As a matter of fact, the enormous corpus of practical experiences and previous studies have pre-eminently revealed that it is difficult to achieve excellent comfortability for the driver and the passengers and simultaneously have a decent road-holding when the car is subjected to exogenous and endogenous perturbations such as different road roughness profiles, an unpredictable variation of vehicle speed and load variation [3-8]. Therefore, the traditional protocol of designing VSS has been and it is still centred on the basis of a trade-off between ride comfort and road holding. Many researchers have attempted to optimise quarter car, half car and full car parameters (QC, HC and FC) models for all types of VSS with the primary objective of probably eliminating the foregoing trade-off [4, 6-9].

Puneet at al., have applied design of experiment (DOE) and multi-objective genetic algorithm (MOGA) to optimise the input QC parameters for better ride comfort and road holding wherein they used a passive damper [6]. Mitra et al., have demonstrated that MOGA proved effective in optimising the QC parameters when PSS is deployed in the vehicle. In comparison with the non-optimised QC parameters, they reported improvement in the ride comfort with a small decrease in road holding value while the non-optimised QCP resulted in significant deterioration of the road holding [4]. When Seifi et al. [7], noticed a critical knowledge gap regarding the foregoing subject matter, they conducted a comprehensive study that presented a method for the optimised design of a PSS in order to improve different aspects of the ride comfort, road holding, workspace and the rollover resistance by considering a full vehicle model with 11-DOF. Similarly, the authors used MOGA to solve their optimisation problem [7]. More recently, Yatak and Sahin have applied a hybrid fuzzy controller to improve the ride comfort-road holding trade-off characteristics of a full VSS [8]. In the work of Anandan and Kandavel, they proposed an ASS composed of a hydraulic actuator and a proportional-integral-derivative controller. Their simulated results alongside MOGA-tuning of the controller parameters showed that the proposed ASS significantly improved the ride comfort with guaranteed vehicle stability [9]. After an extensive literature survey, the authors of the present work saw that a lot of research effort is still needed to upgrade PSS to a level where the previously mentioned trade-off is negligible. As for ASS, they are characterised with satisfactory performance with MOGA optimisation and even better performance when equipped with proportional integral derivative (PID) controllers. In addition, ASS poses an advantage over the PSS to reduce the traditional design as a compromise between vehicle road handling and the occupant riding comfort by directly controlling the suspensions force actuators by using different controllers such as PIDs and linear-quadratic regulator (LQR) control [10,11]. The optimisation of the controller parameters can be performed using different optimisation techniques like a genetic algorithm (GA), particle swarm optimisation (PSO), simulated annealing (SA) technique and other optimisation techniques. The genetic algorithm is selected in this study due to its ability to handle complex problems and proven performance regarding VSS control [6-7].

A genetic algorithm (GA) is the evolutionary optimisation technique that mimics the natural selection process to solve constrained and unconstrained optimisation problems. The algorithm iteratively modifies the initial population of individual solutions. At each step, the GA selects individuals from the current population of solutions and considers them as parents to apply genetic operations to produce the children for the next generation. Over successive generations, the population "evolves" and the optimal solution is reached. The GA performs well where other optimisation techniques fail and are not well suited, which includes the problems in which the objective function is nondifferentiable, stochastic, discontinuous, or highly non-linear [12]. The GA differs from a classical, derivative-based, optimisation algorithm in a sense that while a classical algorithm generates a single solution at each iteration for the sequence to approach an optimal solution, the GA generates a population of solutions at each generation for the population to approach an optimal solution. Also, while the classical approach selects the next point in the sequence of solutions by a deterministic computation, the GA selects the next population by randomised operations [13]. This is an indication that the GA is well suited for the present study, which is also pre-eminently corroborated by the works of the authors in the preceding paragraph [6-10].

Concerning PID control, it is widely used in industrial control systems which employ the control loop mechanism through feedback and a variety of other applications where continuously modulated control is essential. A PID controller uses the error term calculated continuously from the difference of a measured process variable and the desired setpoint or reference of that variable and applies a corrective action based on proportional, integral, and derivative terms. The industrial control applications of the PID control include the regulation temperature, flow, pressure, speed and other process variables [14]. The PID control is a control strategy of choice because it is easy to implement and is a well-established way of driving the desired system towards a target position or level [15]. Ahmed et al., have been able to establish that in a quarter car model, the introduction of an active element in the VSS with a PID controller can achieve a better trade-off than is possible using purely passive elements [15]. Nagarkar et al., demonstrated that MOGA-optimized QC parameters in conjunction with PID control of the ASS gave better control and ride comfort in comparison with classic parameters and PSS [11]. Tandel et al., have also found that implementation of PID control decreases the body acceleration of ASS to the almost half of PSS and therefore the ride comfort of the driver and passengers can be improved. One can agree that the GA optimisation is a viable option to tune PID parameters for enhanced performance of ASS. The proposed approach is tested for the optimal vehicle suspension design meeting the following basic requirements: ride comfort, reduction of dynamic road-tire forces, and the reduction of relative motions between the vehicle bodies [16].

Hence, the primary objective of this study is to design the suspension system to achieve the balance between rider comfort and road handling. In this study, two different models with varied complexities are used to model the suspension system. Quarter and full car models with an increasing level of complexity were used to carry the modal simulations. The state-space representation of the equations of motion considered during this study makes it easy to use the MATLAB program to obtain the response of the system. The objective function of the GA was to minimise various performance responses such as the sprung mass displacement, unsprung mass displacement and the suspension travel. The modelling, simulation and optimisation were done with the aid of MATLAB computational resources. The choice of this computational tool was guided by the successful historical records of their implementation and performance with respect to similar suspension problems [14-16].

## 2. Vehicle Dynamics

The vehicle dynamics (VD) is a part of engineering primarily concerned with the mechanics of vehicle systems. The VD plays a key role in the development of the automobile industry, being a fundamental theory of the industry. The VD for a car can be defined as the response of the vehicle to a driver's inputs on a given solid surface [17]. The VD can be subdivided into longitudinal, lateral and vertical VD. The longitudinal and lateral aspects of the vehicle motion are part of first two sections which include driving, braking and cornering, while the latter concerns with the vertical motion of the vehicle and intern related to the suspension system and rider comfort. The essential parts of the vertical VD system usually comprise the body of the vehicle (sprung mass), the suspension component (spring and damper) and the tire (unsprung mass). The ride comfort focuses on vehicle vibration and pitched movement caused by vertical tire force. Hence, in this study, the vertical dynamics was adapted as it suitable for the objective of the study [18]. These models include quarter car (QC), half car (HC) and the full car (FC) models with an increasing level of complexity and ability to capture details of the dynamics of the vehicle [17-20]. Only the QC and FC models are pertinent to this study.

### 2.1. Quarter Car Model

In the case of QC model, the mass of the vehicle body and the occupants is represented as a single mass element called the sprung mass ($M_s$). The unsprung mass ($M_u$) consists of the mass of the wheel assembly. A spring and a damper are used to model the suspension system between the sprung and unsprung masses while another set of spring and damper are used to model the tire stiffness and internal damping respectively. For the QC representation of ASS, only the actuating force is considered. The actuator is not modelled explicitly for simplicity, and hence the actuator dynamics are not considered making it an ideal actuator. As QC model is most simple to implement, it comes with the inherent limitations. The roll and pitch motion of the vehicle cannot be modelled using QC model. Also, the effect of dependent motion of individual wheel suspensions is not taken into consideration. Fig.1 shows the

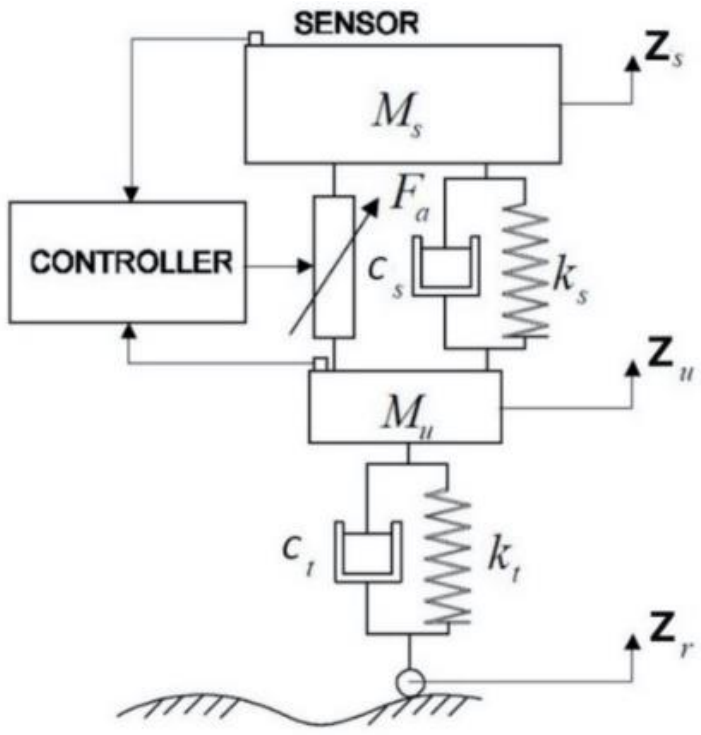

**Figure 1:** Schematic of a quarter car model [17]

The equations of motion for the active QC model are given in Eq. (2.1) where $Z_s$ and $Z_u$ are the vertical displacements of the sprung and unsprung mass, respectively. The road profile is denoted as $Z_r$. The derivatives, i.e., the velocity and accelerations of the sprung and unsprung masses, are represented by a single and double dot over the variables, respectively. $k_s$ is the stiffness of the suspension spring and $C_s$ is the damping coefficient of the suspension damper. The tire stiffness is represented by $k_t$ and internal damping coefficient of the tire is $C_t$. The actuation force is given as $F_a$. The QC model for the PSS system can be obtained simply by removing the actuation force term from the following equations.

$$\begin{aligned} M_s\ddot{Z}_s &= -C_s(\dot{Z}_s - \dot{Z}_u) - k_s(Z_s - Z_u) + F_a \\ M_u\ddot{Z}_u &= C_s(\dot{Z}_s - \dot{Z}_u) + k_s(Z_s - Z_u) - C_t(\dot{Z}_u - \dot{Z}_r) - k_t(Z_u - Z_r) - F_a \end{aligned} \tag{2.1}$$

The standard state-space model equations are given as eq. (2.2).

$$\begin{aligned} \dot{X}(t) &= A\,X(t) + B\,U(t) \\ Y(t) &= C\,X(t) + D\,U(t) \end{aligned} \tag{2.2}$$

where, $A$ = State space matrix, $B$ = Input matrix, $C$ = Output matrix, $D$ = Direct transmission matrix, $X$ = State variables, $U$ = Input of system, $Y$ = Output of system.

It is important to note that State Space Model (SSM) refers to a class of model that describes the dependence between the latent state variable and the observed measurement. The state or the measurement can be either continuous or discrete. In control engineering, a state-space representation is a mathematical model of a physical system as a set of input, output and state variables related by first-order differential equations or difference equations [21]. The state variables are selected ($X(t)$) based on the desired outputs as given in Eq. (2.3).

$$
\begin{aligned}
x_1 &= Z_s \\
x_2 &= \dot{x_1} = \dot{Z}_s \\
x_3 &= Z_u \\
x_4 &= \dot{x_3} = \dot{Z}_u \\
x_5 &= Z_r
\end{aligned} \qquad (2.3)
$$

The left-hand side of the first state-space equation is obtained by taking the first derivative of the state variable vector to be

$$
(X(t) = [x_1 \; x_2 \; x_3 \; x_4 \; x_5]^T) \Rightarrow \dot{X}(t) = [\dot{x_1} \; \dot{x_2} \; \dot{x_3} \; \dot{x_4} \; \dot{x_5}]^T \qquad (2.4)
$$

Now substituting the state variables in the equations of motion to get the desired values of first derivatives of state variables:

$$
\begin{aligned}
M_s\dot{x_2} &= -C_s(x_2 - x_4) - k_s(x_1 - x_3) + F_a \\
\Rightarrow \dot{x_2} &= \frac{1}{M_s}[-C_s(x_2 - x_4) - k_s(x_1 - x_3) + F_a] \\
M_u\dot{x_4} &= C_s(x_2 - x_4) + k_s(x_1 - x_3) - C_t(x_4 - \dot{Z}_r) - k_t(x_3 - x_5) \\
\Rightarrow \dot{x_4} &= \frac{1}{M_u}[C_s(x_2 - x_4) + k_s(x_1 - x_3) - C_t(x_4 - \dot{Z}_r) - k_t(x_3 - x_5) - F_a]
\end{aligned} \qquad (2.5)
$$

By rearranging the above equations, the representations of the first derivatives of the state variables are obtained.

$$
\begin{aligned}
\dot{x_1} &= x_2 \\
\dot{x_2} &= -\frac{k_s}{M_s}x_1 - \frac{C_s}{M_s}x_2 + \frac{k_s}{M_s}x_3 + \frac{C_s}{M_s}x_4 + \frac{1}{M_s}F_a \\
\dot{x_3} &= x_4 \\
\dot{x_4} &= \frac{k_s}{M_u}x_1 + \frac{C_s}{M_u}x_2 - \frac{(k_s + k_t)}{M_u}x_3 - \frac{(C_s + C_t)}{M_u}x_4 + \frac{k_t}{M_u}x_5 + \frac{C_t}{M_u}\dot{Z}_r - \frac{1}{M_u}F_a \\
\dot{x_5} &= \dot{Z}_r
\end{aligned} \qquad (2.6)
$$

By writing the above equations in the matrix form, the following state-space representation is obtained.

$$
\begin{bmatrix} \dot{x_1} \\ \dot{x_2} \\ \dot{x_3} \\ \dot{x_4} \\ \dot{x_5} \end{bmatrix} = \begin{bmatrix} 0 & 1 & 0 & 0 & 0 \\ -\frac{k_s}{M_s} & -\frac{C_s}{M_s} & \frac{k_s}{M_s} & \frac{C_s}{M_s} & 0 \\ 0 & 0 & 0 & 1 & 0 \\ \frac{k_s}{M_u} & \frac{C_s}{M_u} & -\frac{(k_s + k_t)}{M_u} & -\frac{(C_s + C_t)}{M_u} & \frac{k_t}{M_u} \\ 0 & 0 & 0 & 0 & 0 \end{bmatrix} \begin{bmatrix} x_1 \\ x_2 \\ x_3 \\ x_4 \\ x_5 \end{bmatrix} + \begin{bmatrix} 0 & 0 \\ 0 & \frac{1}{M_s} \\ 0 & 0 \\ \frac{C_t}{M_u} & -\frac{1}{M_u} \\ 1 & 0 \end{bmatrix} \begin{bmatrix} \dot{Z}_r \\ F_a \end{bmatrix} \qquad (2.7)
$$

$$
\begin{bmatrix} x_5 \\ x_1 \\ x_2 \\ x_3 \\ x_4 \end{bmatrix} = \begin{bmatrix} Z_r \\ Z_s \\ \dot{Z}_s \\ Z_u \\ \dot{Z}_u \end{bmatrix} = \begin{bmatrix} 0 & 0 & 0 & 0 & 1 \\ 1 & 0 & 0 & 0 & 0 \\ 0 & 1 & 0 & 0 & 0 \\ 0 & 0 & 1 & 0 & 0 \\ 0 & 0 & 0 & 1 & 0 \\ 1 & 0 & -1 & 0 & 0 \end{bmatrix} \begin{bmatrix} x_1 \\ x_2 \\ x_3 \\ x_4 \\ x_5 \end{bmatrix} + \begin{bmatrix} 0 & 0 \\ 0 & 0 \\ 0 & 0 \\ 0 & 0 \\ 0 & 0 \\ 0 & 0 \end{bmatrix} \begin{bmatrix} \dot{Z}_r \\ F_a \end{bmatrix} \qquad (2.8)
$$

Notice that the input is $\dot{Z}_r$ but the road profile is $Z_r$, which can be obtained by differentiating the input.

## 2.2. Full Car Model

In contrast with QC model, FC is a reasonably complex model and hence can capture the detailed dynamics of the vertical motion of the vehicle. The FC model consists of the sprung mass and four unsprung masses are connected to it at four corners through the spring and damper of the suspension system. The tires are modelled as parallel sets of spring and damping elements. The model considered in the study has seven DOFs, which include the heave, roll and pitch of sprung mass and vertical displacement of four unsprung masses. All the elements of the suspension system and the tire stiffness and damping are assumed to be linear. It is also assumed that the location of the centre of gravity (CG) of the sprung mass does not change with time and the coordinate system of the suspension system is attached to the CG of the vehicle and is aligned with the principal axes of the vehicle body. The roll and pitch angles induced during the operation are assumed to be small, and the small-angle approximation is used to obtain the equations of motion, which also adds to the limitation of the model [17]. Like QC model, only the actuating force is considered, and the actuator is not modelled explicitly for the ASS case. The longitudinal, lateral and yaw DOF and the suspension system affect each other. However, this effect is not considered in this model.

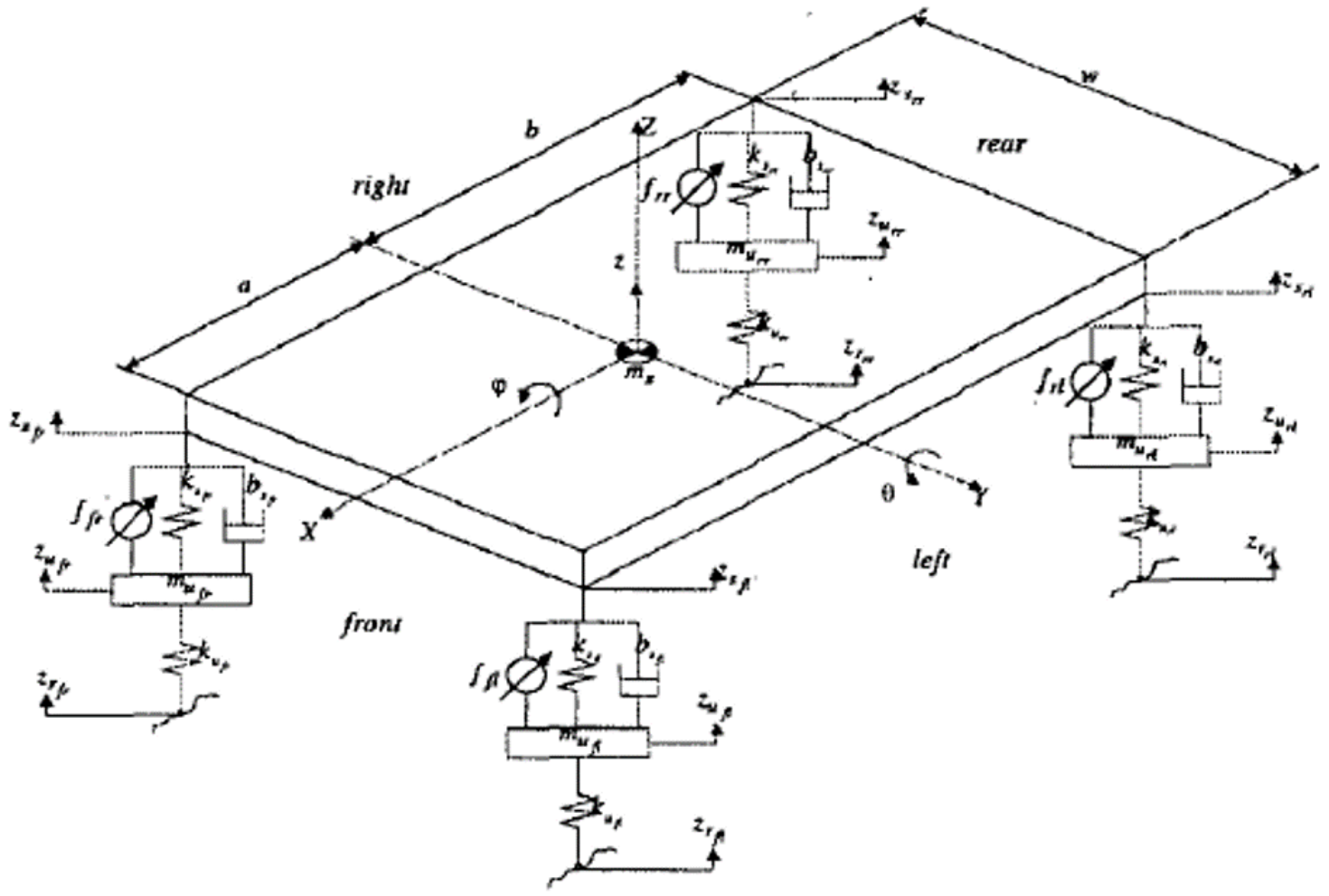

*Figure 2: Schematic of a full car model [17]*

After applying a force-balance analysis to the model in Fig. 2., the equations of motion for the FC model are given as:

$$
\begin{aligned}
M_s\ddot{Z} \quad &= -(2K_{s_f} + 2K_{s_r})Z - (2C_{s_f} + 2C_{s_r})\dot{Z} + (2aK_{s_f} - 2bK_{s_r})\theta + \\
&\quad (2aC_{s_f} - 2bC_{s_r})\dot{\theta} + K_{s_f}Z_{u_{fl}} + C_{s_f}\dot{Z}_{u_{fl}} + K_{s_f}Z_{u_{fr}} + C_{s_f}\dot{Z}_{u_{fr}} + \\
&\quad K_{s_r}Z_{u_{rl}} + C_{s_r}\dot{Z}_{u_{rl}} + K_{s_r}Z_{u_{rr}} + C_{s_r}\dot{Z}_{u_{rr}} \\
I_{yy}\ddot{\theta} \quad &= (2aK_{s_f} - 2bK_{s_r})Z + (2aC_{s_f} - 2bC_{s_r})\dot{Z} - (2a^2K_{s_f} + 2b^2K_{s_r})\theta + \\
&\quad -(2a^2C_{s_f} + 2b^2C_{s_r})\dot{\theta} - aK_{s_f}Z_{u_{fl}} - aC_{s_f}\dot{Z}_{u_{fl}} - aK_{s_f}Z_{u_{fr}} - \\
&\quad aC_{s_f}\dot{Z}_{u_{fr}} + bK_{s_r}Z_{u_{rl}} + bC_{s_r}\dot{Z}_{u_{rl}} + bK_{s_r}Z_{u_{rr}} + bC_{s_r}\dot{Z}_{u_{rr}} \\
I_{xx}\ddot{\phi} \quad &= -0.25w^2(2K_{s_f} + 2K_{s_r})\phi - 0.25w^2(2C_{s_f} + 2C_{s_r})\dot{\phi} + 0.5wK_{s_f}Z_{u_{fl}} + \\
&\quad 0.5wC_{s_f}\dot{Z}_{u_{fl}} - 0.5wK_{s_f}Z_{u_{fr}} - 0.5wC_{s_f}\dot{Z}_{u_{fr}} + 0.5wK_{s_r}Z_{u_{rl}} + \\
&\quad 0.5wC_{s_r}\dot{Z}_{u_{rl}} - 0.5wK_{s_r}Z_{u_{rr}} - 0.5wC_{s_r}\dot{Z}_{u_{rr}} \qquad (2.9)
\end{aligned}
$$

$$
\begin{aligned}
M_u\ddot{Z}_{u_{fl}} \quad &= K_{s_f}Z + C_{s_f}\dot{Z} - aK_{s_f}\theta - aC_{s_f}\dot{\theta} + 0.5wK_{s_f}\phi + 0.5wC_{s_f}\dot{\phi} - \\
&\quad (K_{s_f} + K_t)Z_{u_{fl}} - (C_{s_f} + C_t)\dot{Z}_{u_{fl}} + K_tZ_{r_{fl}} + C_t\dot{Z}_{r_{fl}} \\
M_u\ddot{Z}_{u_{fr}} \quad &= K_{s_f}Z + C_{s_f}\dot{Z} - aK_{s_f}\theta - aC_{s_f}\dot{\theta} - 0.5wK_{s_f}\phi - 0.5wC_{s_f}\dot{\phi} - \\
&\quad (K_{s_f} + K_t)Z_{u_{fr}} - (C_{s_f} + C_t)\dot{Z}_{u_{fr}} + K_tZ_{r_{fr}} + C_t\dot{Z}_{r_{fr}} \\
M_u\ddot{Z}_{u_{rl}} \quad &= K_{s_r}Z + C_{s_r}\dot{Z} + bK_{s_r}\theta + bC_{s_r}\dot{\theta} + 0.5wK_{s_r}\phi + 0.5wC_{s_r}\dot{\phi} - \\
&\quad (K_{s_r} + K_t)Z_{u_{rl}} - (C_{s_r} + C_t)\dot{Z}_{u_{rl}} + K_tZ_{r_{rl}} + C_t\dot{Z}_{r_{rl}} \\
M_u\ddot{Z}_{u_{rr}} \quad &= K_{s_r}Z + C_{s_r}\dot{Z} + bK_{s_r}\theta + bC_{s_r}\dot{\theta} - 0.5wK_{s_r}\phi - 0.5wC_{s_r}\dot{\phi} - \\
&\quad (K_{s_r} + K_t)Z_{u_{rr}} - (C_{s_r} + C_t)\dot{Z}_{u_{rr}} + K_tZ_{r_{rr}} + C_t\dot{Z}_{r_{rr}}
\end{aligned}
$$

Where, $M_s$ and $M_u$ have a similar meaning as QC model and are sprung and unsprung mass, respectively. The additional roll and pitch DOFs come into picture due to the moment of inertia $I_{xx}$ and $I_{yy}$ about the roll and pitch axes of the vehicle respectively. The pitch angle is represented as $\theta$ and the roll angle is represented as $\phi$. Z is the vertical displacement of the sprung mass while the unsprung mass and road profile are denoted as $Z_u$ and $Z_r$ with the subscripts corresponding to the wheel. Here *fr* stands for front-right, *fl* is front-left, *rr* is rear-right and *rl* stands for rear-left. A single dot above the variable denotes the velocity, while the acceleration is given as double dots. The spring stiffness and the damping coefficient are demoted as $K_s$ and $C_s$ while the front and rear variations are denoted with the subscripts *f* and *r*, respectively. The tire stiffness is $K_t$ and the tire internal damping coefficient is $C_t$.

The equations of motion are represented in the form of state-space equations. The selected states are given in eq. (2.10).

$$
\begin{aligned}
x_1 &= Z \\
x_2 &= \dot{Z} \\
x_3 &= \theta \\
x_4 &= \dot{\theta} \\
x_5 &= \phi \\
x_6 &= \dot{\phi} \\
x_7 &= Z_{u_{fl}} \\
x_8 &= \dot{Z}_{u_{fl}} \\
x_9 &= Z_{u_{fr}} \\
x_{10} &= \dot{Z}_{u_{fr}} \\
x_{11} &= Z_{u_{rl}} \\
x_{12} &= \dot{Z}_{u_{rl}} \\
x_{13} &= Z_{u_{rr}} \\
x_{14} &= \dot{Z}_{u_{rr}} \\
x_{15} &= Z_{r_{fl}} \\
x_{16} &= Z_{r_{fr}} \\
x_{17} &= Z_{r_{rl}} \\
x_{18} &= Z_{r_{rr}}
\end{aligned}
\tag{2.10}
$$

The derivatives of the states can be obtained by rearranging the equations of motion.

$$
\begin{aligned}
\dot{x_1} &= x_2 \\
\dot{x_2} &= \frac{1}{Ms}\{-2(K_{s_f} + K_{s_r})x_1 - 2(C_{s_f} + C_{s_r})x_2 + 2(aK_{s_f} - b * K_{s_r})x_3 \\
&\quad +2(aC_{s_f} - b * C_{s_r})x_4 + K_{s_f}x_7 + C_{s_f}x_8 + K_{s_f}x_9 + C_{s_f}x_{10} \\
&\quad +K_{s_r}x_{11} + C_{s_r}x_{12} + K_{s_r}x_{13} + C_{s_r}x_{14}\} \\
\dot{x_3} &= x_4 \\
\dot{x_4} &= \frac{1}{I_{yy}}\{2(aK_{s_f} - bK_{s_r})x_1 + 2(aC_{s_f} - bC_{s_r})x_2 - 2(a^2K_{s_f} + b^2K_{s_r})x_3 \\
&\quad -2(a^2C_{s_f} + b^2C_{s_r})x_4 - aK_{s_f}x_7 - aC_{s_f}x_8 - aK_{s_f}x_9 - aC_{s_f}x_{10} \\
&\quad bK_{s_r}x_{11} + bC_{s_r}x_{12} + bK_{s_r}x_{13} + bC_{s_r}x_{14}\} \\
\dot{x_5} &= x_6 \\
\dot{x_6} &= \frac{1}{2I_{xx}}\{-w^2(K_{s_f} + K_{s_r})x_5 - w^2(C_{s_f} + C_{s_r})x_6 + wK_{s_f}x_7 + wC_{s_f}x_8 \\
&\quad -wK_{s_f}x_9 - wC_{s_f}x_{10} + wK_{s_r}x_{11} + wC_{s_r}x_{12} - wK_{s_r}x_{13} - wC_{s_r}x_{14}\} \\
\dot{x_7} &= x_8 \\
\dot{x_8} &= \frac{1}{Mu}\{K_{s_f}x_1 + C_{s_f}x_2 - aK_{s_f}x_3 - aC_{s_f}x_4 + \frac{w}{2}K_{s_f}x_5 + \frac{w}{2}C_{s_f}x_6 \\
&\quad -(K_{s_f} + K_t)x_7 - (C_{s_f} + C_t)x_8 + K_t x_{15} + C_t\dot{Z}_{r_{fl}}\} \\
\dot{x_9} &= x_{10}
\end{aligned}
\tag{2.11}
$$

$$
\begin{aligned}
\dot{x_{10}} &= \frac{1}{Mu}\{K_{s_f}x_1 + C_{s_f}x_2 - aK_{s_f}x_3 - aC_{s_f}x_4 - \frac{w}{2}K_{s_f}x_5 - \frac{w}{2}C_{s_f}x_6 \\
&\quad -(K_{s_f} + K_t)x_9 - (C_{s_f} + C_t)x_{10} + K_t x_{16} + C_t \dot{Z}_{r_{fr}}\} \\
\dot{x_{11}} &= x_{12} \\
\dot{x_{12}} &= \frac{1}{Mu}\{K_{s_r}x_1 + C_{s_r}x_2 + bK_{s_r}x_3 + bC_{s_r}x_4 + \frac{w}{2}K_{s_r}x_5 + \frac{w}{2}C_{s_r}x_6 \\
&\quad -(K_{s_r} + K_t)x_{11} - (C_{s_r} + C_t)x_{12} + K_t x_{17} + C_t \dot{Z}_{r_{rl}}\} \\
\dot{x_{13}} &= x_{14} \\
\dot{x_{14}} &= \frac{1}{Mu}\{K_{s_r}x_1 + C_{s_r}x_2 - aK_{s_r}x_3 - aC_{s_r}x_4 - \frac{w}{2}K_{s_r}x_5 - \frac{w}{2}C_{s_r}x_6 \\
&\quad -(K_{s_r} + K_t)x_{13} - (C_{s_r} + C_t)x_{14} + K_t x_{18} + C_t \dot{Z}_{r_{rr}}\} \\
\dot{x_{15}} &= \dot{Z}_{r_{fl}} \\
\dot{x_{16}} &= \dot{Z}_{r_{fr}} \\
\dot{x_{17}} &= \dot{Z}_{r_{rl}} \\
\dot{x_{18}} &= \dot{Z}_{r_{rr}}
\end{aligned}
$$

$$
\begin{aligned}
Z_{s_{fl}} &= Z - a\theta + (w/2)\phi \\
Z_{s_{fr}} &= Z - a\theta - (w/2)\phi \\
Z_{s_{rl}} &= Z + b\theta + \left(\frac{w}{2}\right)\phi \qquad (2.12) \\
Z_{s_{rr}} &= Z + b\theta - (w/2)\phi
\end{aligned}
$$

The matrix representation of the above equations is used in the MATLAB implementation.

## 3. Control of the active suspension

The PID control is the most-used feedback control design in control engineering. It shows the three terms on the error signal to produce a control signal. If u(t) is the control signal sent to the system y(t) is the actual output, r(t) is the desired output, and the tracking error e(t) = r(t) – y(t), a PID controller has the next form [21]:

$$u(t) = k_p e(t) + k_i \int e(t)dt + k_d \frac{d}{dt}e(t) \qquad (2.13)$$

The desired closed-loop dynamics can be obtained by adjusting the three parameters Kp, Ki and Kd, often iteratively with "tuning" and without specific knowledge of a plant model. Stability can often be obtained using the proportional term. The integral term permits the rejection of a step disturbance. The derivative term provides damping or shaping of the response. PIDCs are the most well-established class of control systems [21].

The ASS consists of a PID controller which operates on the basis of error signal, which is the suspension travel calculated as the difference in the displacements of sprung and unsprung masses. The PID controller for the QC model is shown in Fig. 3 where $F_a$ and $K$ denote the actuating force and static gain in the system, respectively. As for the FC model, it has four actuators as shown in Fig 4 where each of the four force actuators is having its corresponding PID controller. This translates to the tuning of the total of 12 parameters mentioned elsewhere, 3 for each PID controller. The forces $f_{fr}$, $f_{fl}$, $f_{rr}$ and $f_{rl}$ are the actuating forces and the suspension travels corresponding to each wheel are taken as the error signal.

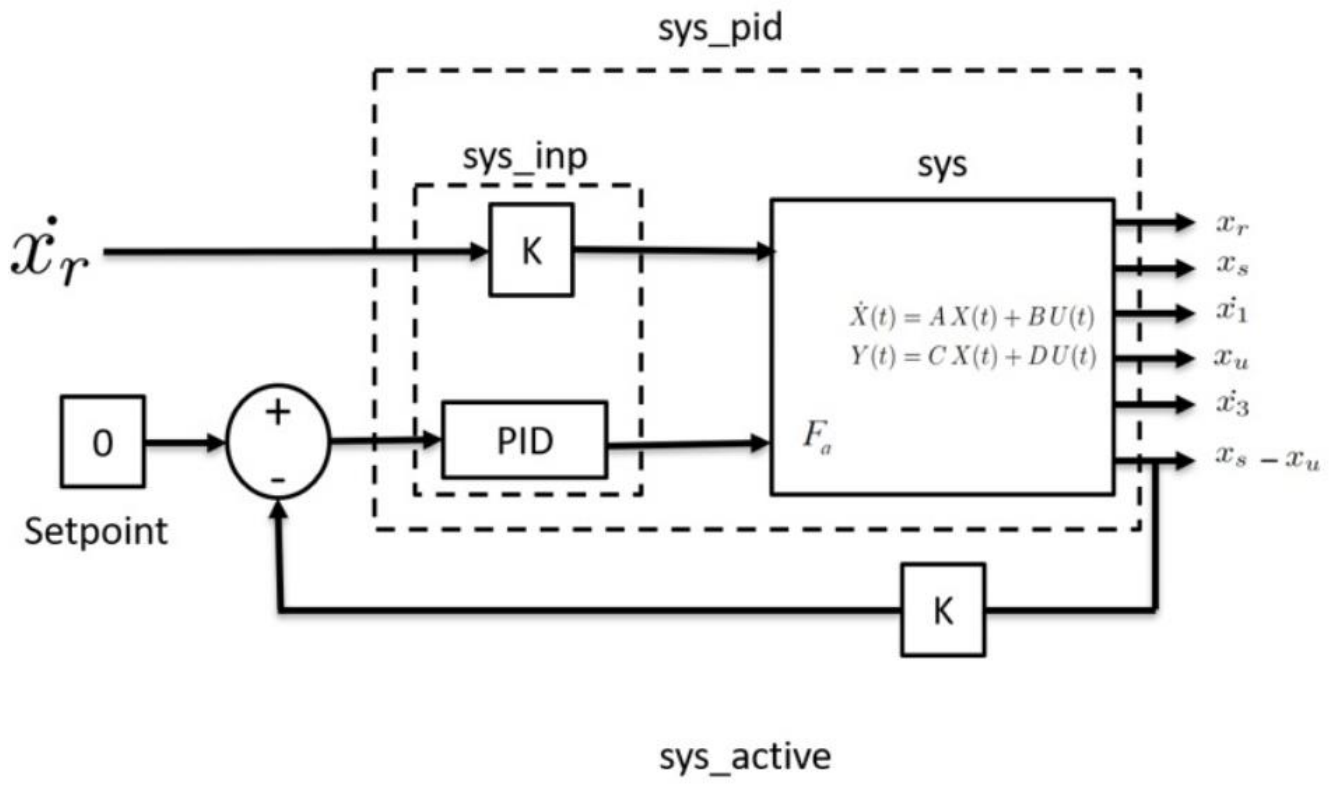

*Figure 3: Block diagram for PID controller for the QC model*

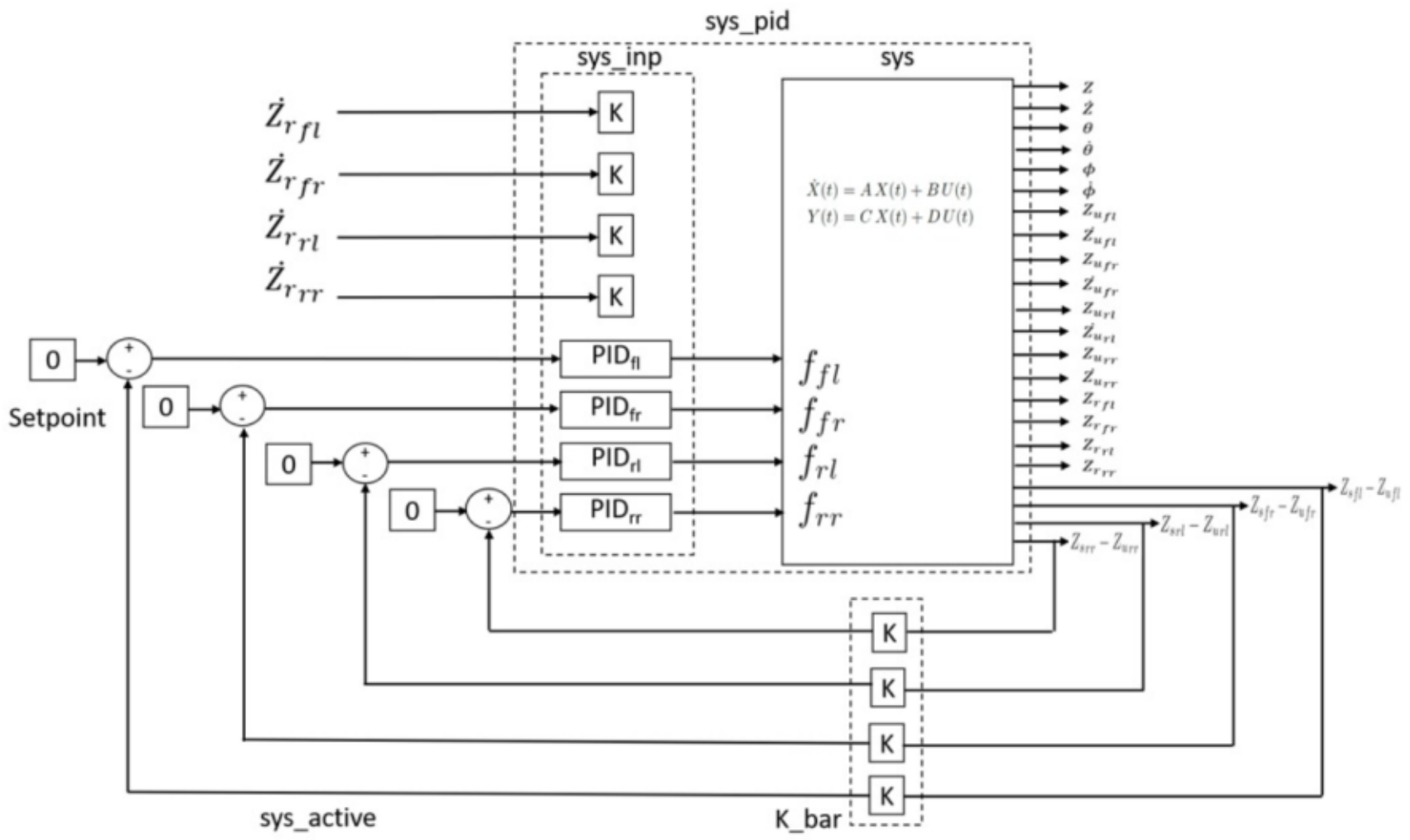

*Figure 4: Block diagram for PID controller for the FC model*

The setpoint is zero in all the cases as the objective is to minimise the suspension travel. Similar to the QC model case, the K is gain with value one included to facilitate the MATLAB implementation.

## 4. Genetic Algorithm Optimisation

GA is an evolutionary algorithm technique to find an optimum solution of the problem, where the problem is affected by a number of factors. This is inspired by biological terms such as mutation, cross-over and reproduction. The GA uses objective functions which in our case are selected as LQR cost function and Constraint-Based (CB) objective functions. Concerning LQR objective function, it is a type of optimal control based state-space representation. It is applicable in a dynamic system described by a set of linear differential equations, and the objective is described by a quadratic function [22]. The penalty matrices, Q and R are the diagonal matrices whose elements are selected manually from the knowledge of the system and the intuition. As for the CB objective function, the objective function is the sum of different physical constraints required to be enforced on the suspension system pertaining to the rider comfort, road handling and vehicle safety[23].

Previous studies have shown that the evolutionary GA and other AI techniques such as fuzzy logic and neural network by virtue of their capabilities to solve multiple-inputs-multiple-outputs optimisation problems are preferable than the traditional PID tuning techniques like Zeigler Nichols, Cohen-coon and pole placement tuning scheme for obtaining optimal gains for PID controllers [24]. The GA protocol is adapted for this due to its flexibility in terms of PID tuning schemes. In this study, the GA-Linear Quadratic Regulator (LQR) optimisation technique was employed for obtaining the optimal PID gains and the GA-Constraint-Based (CB) optimisation technique was employed for obtaining the optimal PID gains and suspension parameters. More importantly, unlike the GA-LQR approach, the GA-CB approach does not involve the onerous process of manually selecting a number of penalties for its implementation [22,23] and only one penalty variable alpha is used in GA-CB. The parameters used for the GA optimisations are illustrated in Table 1.

*Table 1: GA parameters used for both the LQR and CB cases.*

| **Parameter** | **Value** |
|---|---|
| Population size | 100 |
| Number of generations | 50 |
| Constraint Tolerance | 1e-3 |
| Cross-over approach | Scattered |
| Cross-over fraction | 0.8 |
| Elite count | 5 |
| Mutation function | Gaussian |
| Lower bound: [$K_s$ $C_s$ $K_p$ $K_i$ $K_d$ ] | [15000 400 1 1 1] |
| Upper bound: [$K_s$ $C_s$ $K_p$ $K_i$ $K_d$ ] | [80000 5500 150000 150000 150000] |
| Alpha | 10000 |

### 4.1. GA-LQR optimisation of PID controller gains for the ASS

Here, we have taken the cost function from the Linear Quadratic Regulator (LQR) control as the objective function for the GA. The GA uses the objective function from the LQR controller system to arrive at the optimal set of PID parameters. Thus, for a given state-space model;

$$\dot{x} = Ax + Bu$$

The LQR cost function is given as

$$J = \frac{1}{2}\int_0^\infty [x^t Qx + u^T Ru]dt \qquad (2.13)$$

Here, x and u are states and inputs respectively. The Q matrix is a diagonal matrix and the diagonal element are the penalties for the corresponding state-variables. In the case of QC model, the R matrix is the penalty for actuating force $F_a$. Similarly, in the case of FC model, the $R$ matrix is the penalty for actuating force $f_{fl}$, $f_{fr}$, $f_{rl}$ and $f_{rr}$. These penalties are selected manually according to the requirement and understanding of the physical behaviour of the system. The values of these penalties vary from 10 to 100000 to drive responses to the desired performance.

## 4.2. GA-CB Optimization of PID Parameters for ASS in QC model

The next objective is to find the optimum values for spring stiffness ($k_s$) and damping coefficient ($C_s$) for achieving the maximum rider comfort with the quarter car model. The Genetic Algorithm is used to find these optimum parameters. The lower and upper bounds for the optimisation process are selected based on the range of values used in the literature.

The cost function is selected such that the main objective is to achieve the maximum rider comfort. The rider comfort is assessed by the acceleration of the sprung mass ($\ddot{Z}_s$). The minimisation of the RMS (root mean square) of the absolute value of the acceleration for time T seconds is the objective function $f$. There are multiple constraints of this objective function denoted as $g1$ to $g8$.

$$f = RMS|\ddot{Z}_s|$$

The constraints are also included in the cost function **J** to convert the problem into an unconstrained one. These constraints are:

- The ISO standard ISO2631 states that the passenger feels highly comfortable for weighted RMS acceleration maintained below 0.315 m/s$^2$. This is the first constraint.

$$g1 = f - 0.315 m/s^2 \leq 0$$

- To absorb a bump acceleration of "0.5g" without hitting the suspension stops, at least 12.7 cm (5 inches) of suspension travel must be available. Also, the maximum acceleration value is kept under the threshold to avoid hitting the suspension stops during operation. These form the next two constraints.

$$\begin{aligned} g2 &= |Z_s - Z_u| - 0.127m \leq 0 \\ g3 &= max|\ddot{Z}_s| - 4.5m/s^2 \leq 0 \end{aligned}$$

- The dynamic tyre forces are constraints using the maximum tyre deflection.

$$g4 = |Z_u - Z_r| - 0.0508m \leq 0$$

- The other side of the passenger comfort is road handling, which is included in the equation as a constraint.

$$g5 = |Z_u| - 0.07m \leq 0$$

The objective function and the constraints are the same as the passive case. However, three additional constraints are added to incorporate the comfortable frequency range of 0.8 Hz and 1.5 Hz. Also, the jerk experienced by the passengers is considered as a constraint.

$$\begin{aligned} g6 &= 0.8 \leq Wn \leq 1.5Hz \\ g7 &= \left|\dddot{Z}_s\right| - 18m/s^3 \end{aligned}$$

The actuating force should be below a specific limit:

$$g8 = |F_a| - 400N \leq 0$$

In the sixth constraint, the natural frequency is calculated using the suspension parameters using:

$$\text{Ride Rate} = RR = k_s \times k_t/(k_s + k_t)$$

$$\text{Ride Frequency} = f_n = \sqrt{\frac{RR}{M_s}} \times \frac{1}{2\pi} Hz$$

$$\text{Damped Frequency} = f_d = f_n\sqrt{1-\zeta^2} Hz$$

$$\text{Damping Ratio} = \zeta = \frac{C_s}{\sqrt{4k_sM_s}}$$

The total cost function **J** is given by

$$J = f + \alpha \sum_{i=1}^{8} \max(0, g_i)$$

Where $\alpha$ is the penalty for each constraint violation.

### *4.3.* GA-CB Optimization of PID Parameters for ASS in the FC model

In this section, the objective is to find the optimum values for spring stiffnesses for the front ($K_{s_f}$) and rear ($K_{s_r}$) suspension and the damping coefficient for the front ($C_{s_f}$) and rear ($C_{s_r}$) suspension for achieving the maximum rider comfort with the full car model. The PID parameters $K_p$, $K_i$ and $K_d$ are optimised to have optimum actuating force.

The objective function and the constraints are the same as the previous section with few added constraints. The multiple constraints of this objective function are denoted as $g1$ to $g24$.

The constraints are also included in the cost function **J** to convert the problem into an unconstrained one. These constraints are:

$$g1 = f - 0.315m/s^2 \leq 0$$

$$g2 = |Z_{s_{fl}} - Z_{u_{fl}}| - 0.127m \leq 0$$
$$g3 = |Z_{s_{fr}} - Z_{u_{fr}}| - 0.127m \leq 0$$
$$g4 = |Z_{s_{rl}} - Z_{u_{rl}}| - 0.127m \leq 0$$
$$g5 = |Z_{s_{rr}} - Z_{u_{rr}}| - 0.127m \leq 0$$
$$g6 = max|\ddot{Z}| - 4.5m/s^2 \leq 0$$

$$g11 = |Z_{u_{fl}}| - 0.07m \leq 0$$
$$g12 = |Z_{u_{fr}}| - 0.07m \leq 0$$
$$g13 = |Z_{u_{rl}}| - 0.07m \leq 0$$
$$g14 = |Z_{u_{rr}}| - 0.07m \leq 0$$

The front and rear frequency, along with the jerk constraints, are the added constraints.

$$g15 \quad = 0.8 \leq f_{d_f} \leq 1.5Hz$$
$$g16 \quad = 0.8 \leq f_{d_r} \leq 1.5Hz$$
$$g17 \quad = |\ddot{Z}| - 18$$

The natural frequency of front suspension should be greater than the rear suspension

$$g18 \quad = f_{d_f} > f_{d_r}$$

Pitch frequency and roll frequency range should be the same as the ride frequency

$$g19 \quad = 0.8 \leq f_{d_f} \leq 1.5Hz$$
$$g20 \quad = 0.8 \leq f_{d_r} \leq 1.5Hz$$

The actuating force should not exceed a specific range

$$g21 \quad = |F_{a_{fl}}| - 1000N \leq 0$$
$$g22 \quad = |F_{a_{fr}}| - 1000N \leq 0$$
$$g23 \quad = |F_{a_{rl}}| - 1500N \leq 0$$
$$g24 \quad = |F_{a_{rr}}| - 1500N \leq 0$$

The total cost function **J** is given by

$$J \quad = f + G_c$$

Where $G_c$ is the combined effect of the constraints given by

$$G_c \quad = \alpha \times \sum_{i=1}^{24} m\,ax(0, g_i)$$

Where $\alpha$ is a penalty value which will vary between 8000 and 10000.

### 4.4. Road Profile

The road profile is of paramount importance in the understanding of a vehicle's response to endogenous and exogenous perturbations or road excitations. Examples of such perturbations can be encountered when a vehicle meet bumps and potholes. Bumps were selected for investigation in this study. It has been established that, with rectangular cleats, the dynamic reaction of a vehicle or a single tyre to sudden impact can be investigated. If the shape of the obstacle is approximated by a smooth function, like a cosine wave, then discontinuities will be avoided. Usually, deterministic obstacles are described in local coordinate systems as furnished in Figure 5 [24].

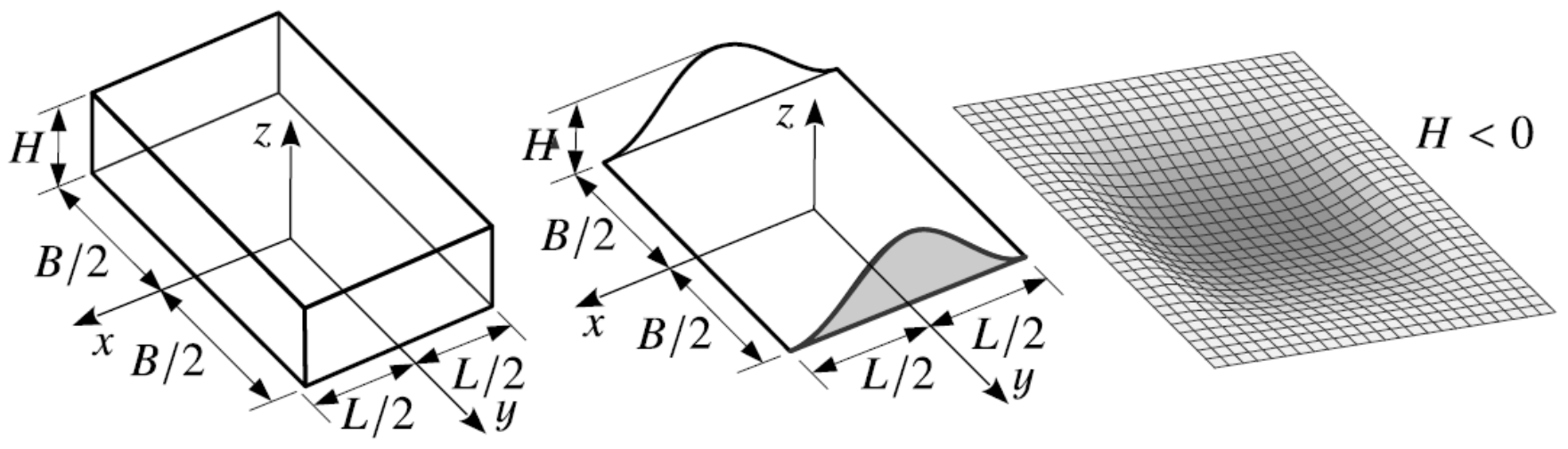

*Figure 5: Rectangular cleat (left), cosine-shape bump (centre), and rounded obstacle or pothole (right) [24]*

Then, the rectangular cleat is simply defined by

$$z(x,y)=\begin{cases}H & \text{if else}, \; -\frac{L}{2}<X<\frac{L}{2} \;\&-\frac{B}{2}<y<\frac{B}{2}\\ 0\end{cases} \qquad (2.14)$$

And the cosine-shaped bump is given by

$$z(x,y)=\begin{cases}\frac{1}{2}H\left(1+\cos\left(2\pi\frac{x}{L}\right)\right) & \text{if else}, \; -\frac{L}{2}<x<\frac{L}{2} \;\&\; -\frac{B}{2}<y<\frac{B}{2}\\ 0\end{cases} \qquad (2.15)$$

Where L, B, and H denote the length, width and height of the obstacle. The cosine-shaped bump was selected for this study [24]. The parameters used for the car models are presented in Table 2.

*Table 1: Parameters used for quarter car and full car models*

| Parameter | Quarter Car | | Full Car | |
|---|---|---|---|---|
| Sprung Mass (kg) | $M_s$ | 375 | $M_s$ | 1500 |
| Unsprung Mass (kg) | $M_u$ | 59 | $M_u$ | 59 |
| Spring stiffness for front suspension (N/m) | $K_s$ | 35000 | $K_{sf}$ | 35000 |
| Spring stiffness for rear suspension (N/m) | - | - | $K_{sr}$ | 3800 |
| Damping coefficient for front suspension (N/m/s) | $C_s$ | 1000 | $C_{sf}$ | 1000 |
| Damping coefficient for rear suspension (N/m/s) | - | - | $C_{sr}$ | 1100 |
| Tire stiffness (N/m) | $K_t$ | 190000 | $K_t$ | 190000 |
| Damping coefficient for tire (N/m/s) | $C_t$ | 2 | $C_t$ | 2 |
| Moment of inertia about roll axis (Kg-m$^2$) | - | - | $I_{xx}$ | 460 |
| Moment of inertia about pitch axis (Kg-m$^2$) | - | - | $I_{yy}$ | 2100 |
| Distance of CG of sprung mass from front axle (m) | - | - | A | 1.4 |
| Distance of CG of sprung mass from front axle (m) | - | - | b | 1.7 |
| Width of sprung mass (m) | - | - | w | 3 |

## 5. Results and Discussion

The simulated results for the PSS and those from the GA protocols for the ASS are presented in Figures 6-9 for the QC model and the FC model. To evaluate the performances of the designed ASS, the results are compared with that of the PSS. The parameters selected for comparisons in the QC model are the sprung mass displacement, unsprung mass displacement, and the suspension travel in the time domain. While for FC model, the pitch angle and the roll angle were added to the parameters. For the QC model, only the PID gains were optimised while the PID gains and other performance parameters were optimised for the FC model. The simulations were performed using the car model parameters given in Table 2, for the desired road profiles and the optimal suspension parameters are presented in Table 3.

### 5.1. Quarter Car Model with GA-LQR Optimization for the PID Gains

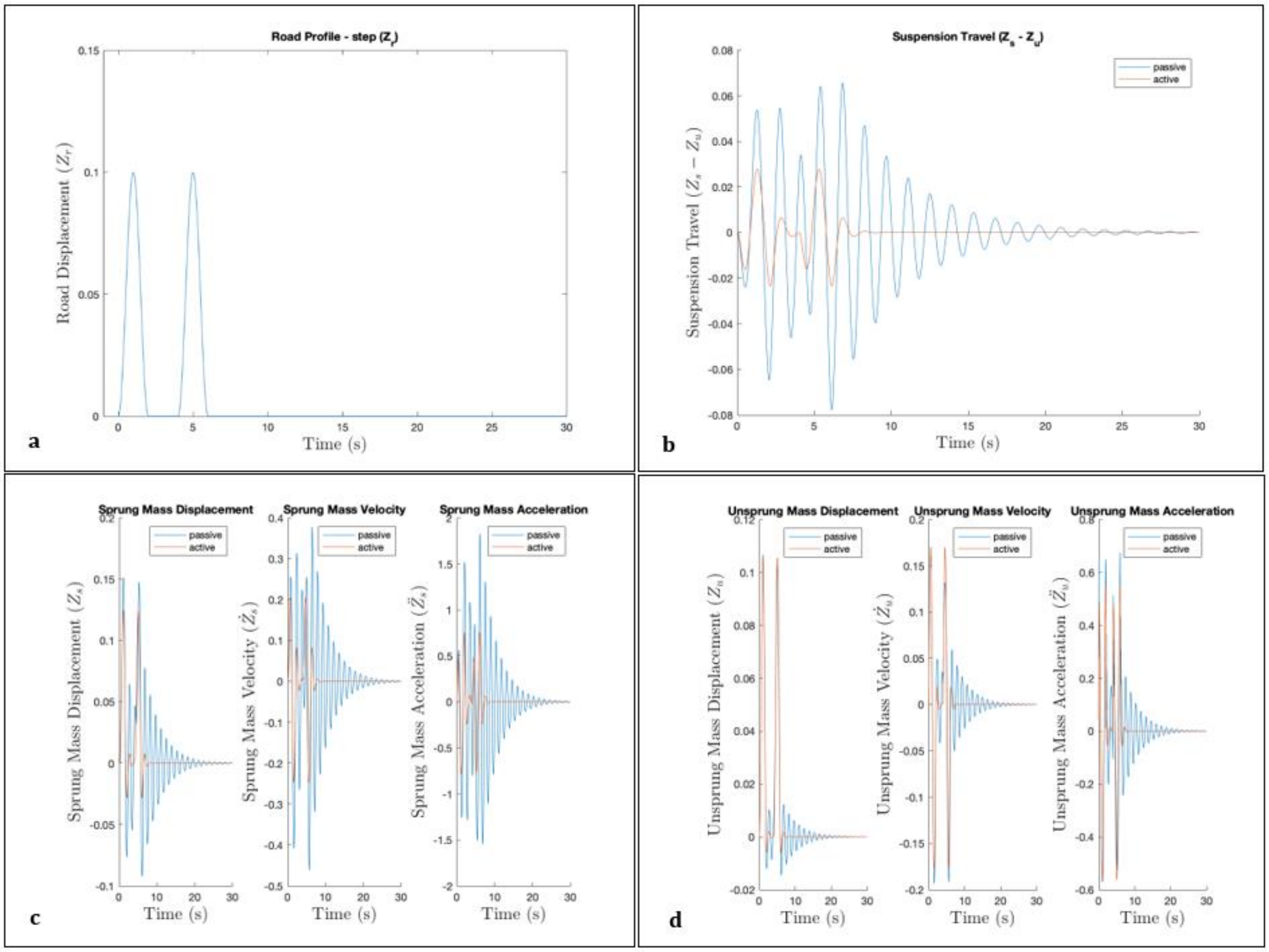

*Figure 6: Plots of (a) road profile versus time (b) suspension vs travel time (c) sprung mass displacement parameters vs time and (d) unsprung mass displacement parameters vs time for the QC model*

Figure 6 (a) shows the road profile with two bumps of 0.1 m, i.e., 10 cm height and Figure 6 (b-d) shows the comparisons of suspension travel, sprung mass displacement, and unsprung mass displacement for the GA-LQR optimised ASS and the PSS in the time domain. It indicates that the suspension travel, the sprung mass displacement and the unsprung mass displacement all reduced with the implementation of the active elements despite the same road profiles. It was observed for all simulated instances that the GA-LQR optimised ASS for the QC model outperformed the PSS. The peak amplitudes and vibration settling time (VST) of the PSS are higher than that of the ASS which translates to improved ride comfort (sprung mass acceleration), road holding (suspension travel) and tyre deflection (unsprung mass acceleration) for the ASS. The VST reduced from around 25 to 7 seconds for the optimised ASS. Similarly, the peak amplitude of sprung mass displacement observed for the optimised ASS is around 17% less than the PSS. The maximum suspension travel for the PSS is almost double of that for the ASS. Nonetheless, both suspension systems settled within the desired travel time. This is an indication that the LQR objective function can be successfully integrated with GA for enhanced performance of an ASS equipped with a PID controller. Mitra et al., and Ahmed et al., have reported similar findings when they compared PSS and ASS in their studies [4,15].

### 5.2. Quarter Car Model with GA-CB Optimization for the PID Gains

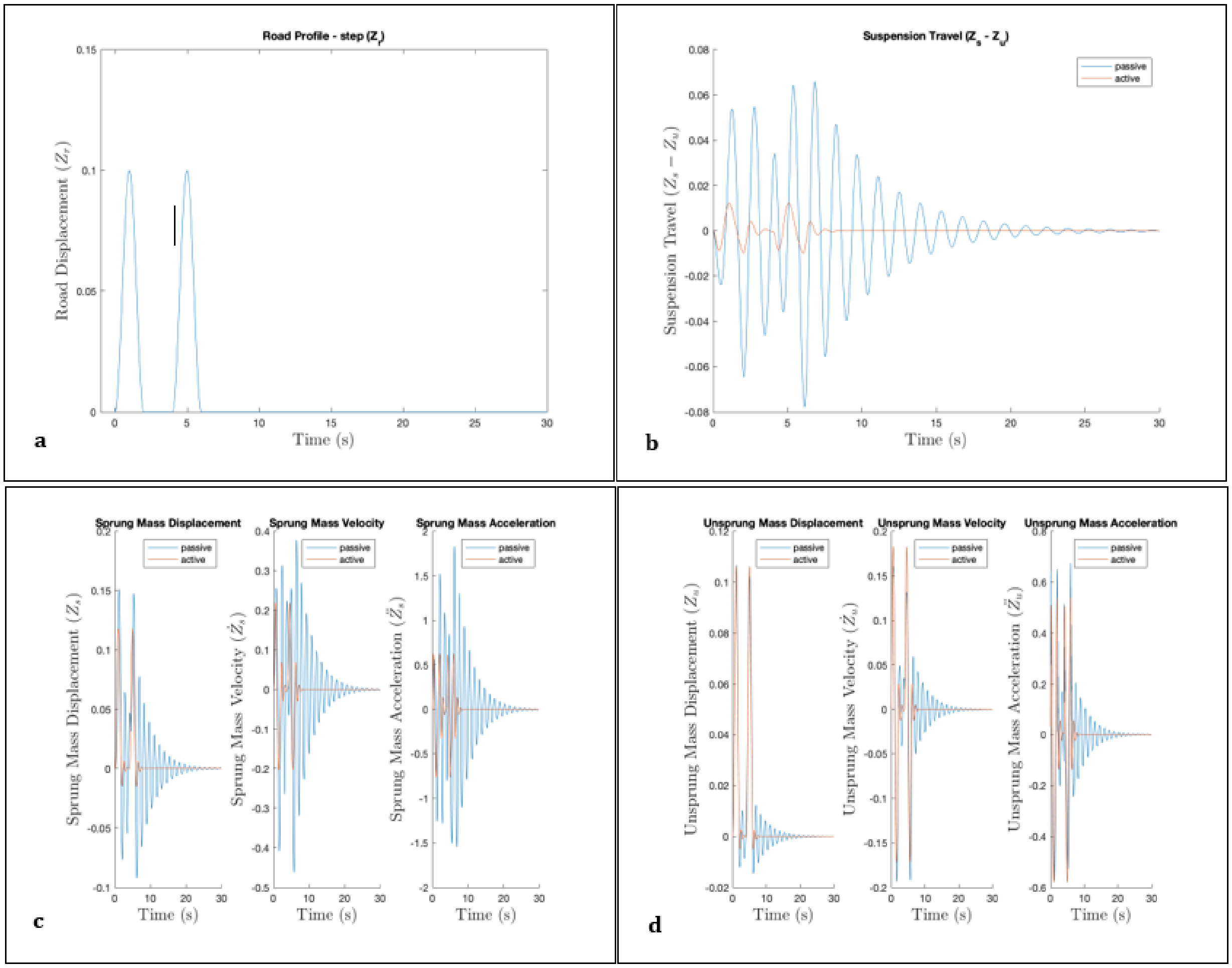

*Figure 7: Plots of (a) road profile versus time (b) suspension vs travel time (c) sprung mass displacement parameters vs time and (d) unsprung mass displacement parameters vs time for the QC model*

Figure 7 (a) shows the road profile with two bumps of 0.1 m, i.e., 10 cm height and Figure 7 (b)-(d) shows the comparisons of suspension travel, sprung mass displacement, and unsprung mass displacement for the GA-LQR optimised ASS and the PSS in the time domain. It indicates that the suspension travel, the sprung mass displacement and the unsprung mass displacement all reduced with the implementation of the active elements despite same road profiles. Correspondingly, the GA-CB optimisation showed similar results in all simulated instances as was observed for the GA-LQR approach. In both the approaches the physical constraints mentioned in the GA-CB sections are followed and the values of maximum suspension travel is less than 0.127m (5 inches) as advised and stayed at the nominal value of around 0.02m. Also, the maximum value of the sprung mass acceleration observed was less than 4.5m/s$^2$ and was below 1 m/s$^2$. Similar to the GA-LQR optimised ASS case, the VST is reduced from 25 to 7 seconds as compared to the PSS. The prominent improvement in the rider comfort can be observed considering the VST. The peak amplitudes of the sprung mass displacement, i.e., the displacement felt by the occupants of the vehicle, is reduced considerably for the optimised ASS. The results presented indicate that there is only a slight difference in the performance of the GA-LQR and GA-CB optimization approach. However, the GA-CB optimisation approach is a better alternative as its implementation involves the use of actual vehicle constraint and a smaller number of manual tunning parameters. Table 3 presents the optimised valued of PID gains for two different optimisation approaches.

*Table 3: Optimised PID gains of GA-LQR and GA-CB approaches for quarter car model*

| PID Gain | GA-LQR | GA-CB |
|---|---|---|
| Kp | 227.13 | 12225 |
| Ki | 1.20 | 22241 |
| Kd | 5878.56 | 841.7 |

## 5.3. Full Car Model with GA-LQR Optimization for the PID Parameters

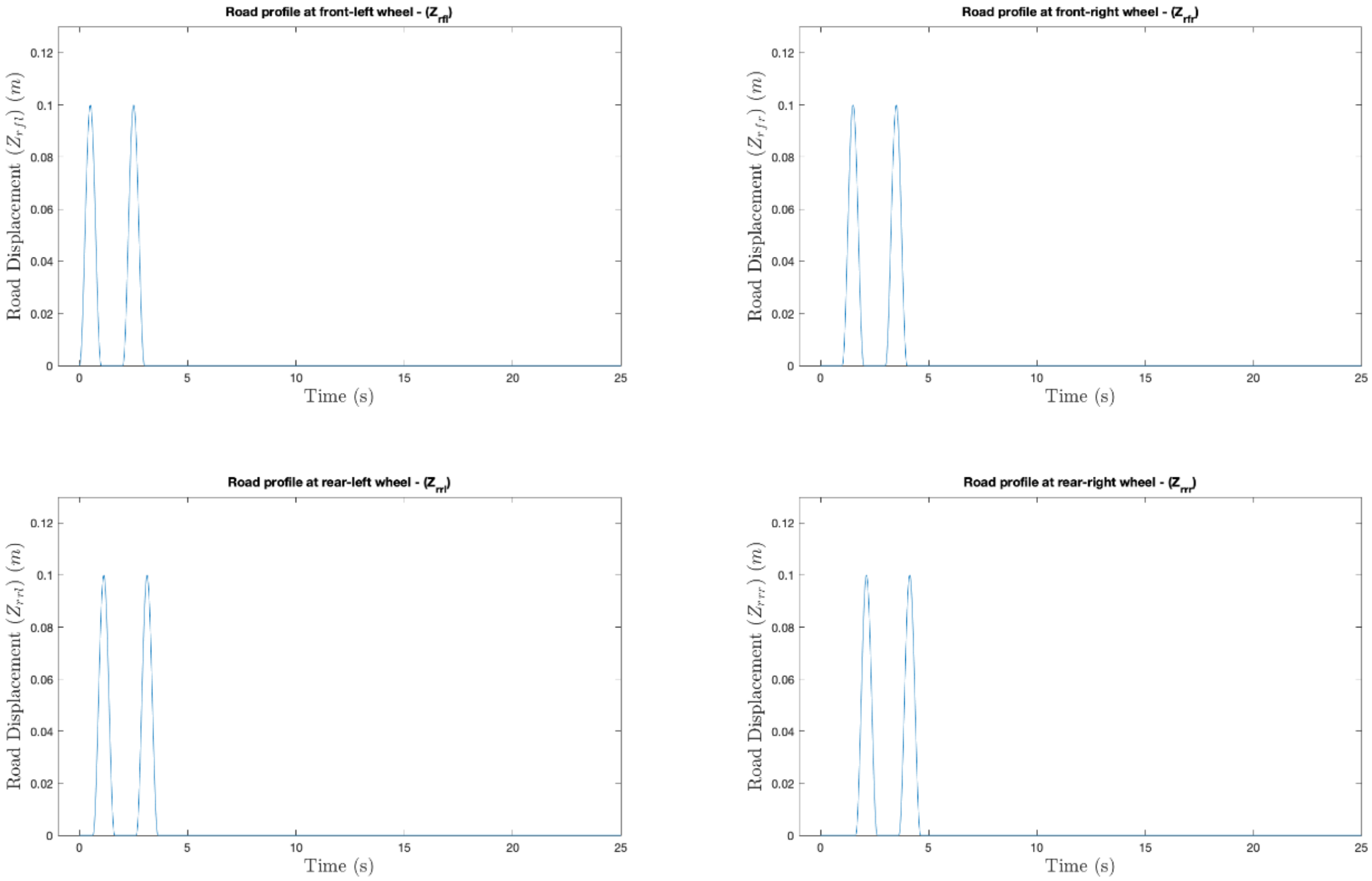

*Figure 8 (a): Plots of road displacement vs time for the FC model (the four wheels are illustrated in the plots)*

Figure 8(a) shows the double bumps of 10 cm height experiences by four wheels of the vehicle. The delay is maintained between the bumps on the right and left side of the vehicle to induce rolling movement in the vehicle. The bumps experienced by the rear wheels also have some delay. However, this delay depends on the wheelbase of the vehicle and the longitudinal speed with which it is travelling.

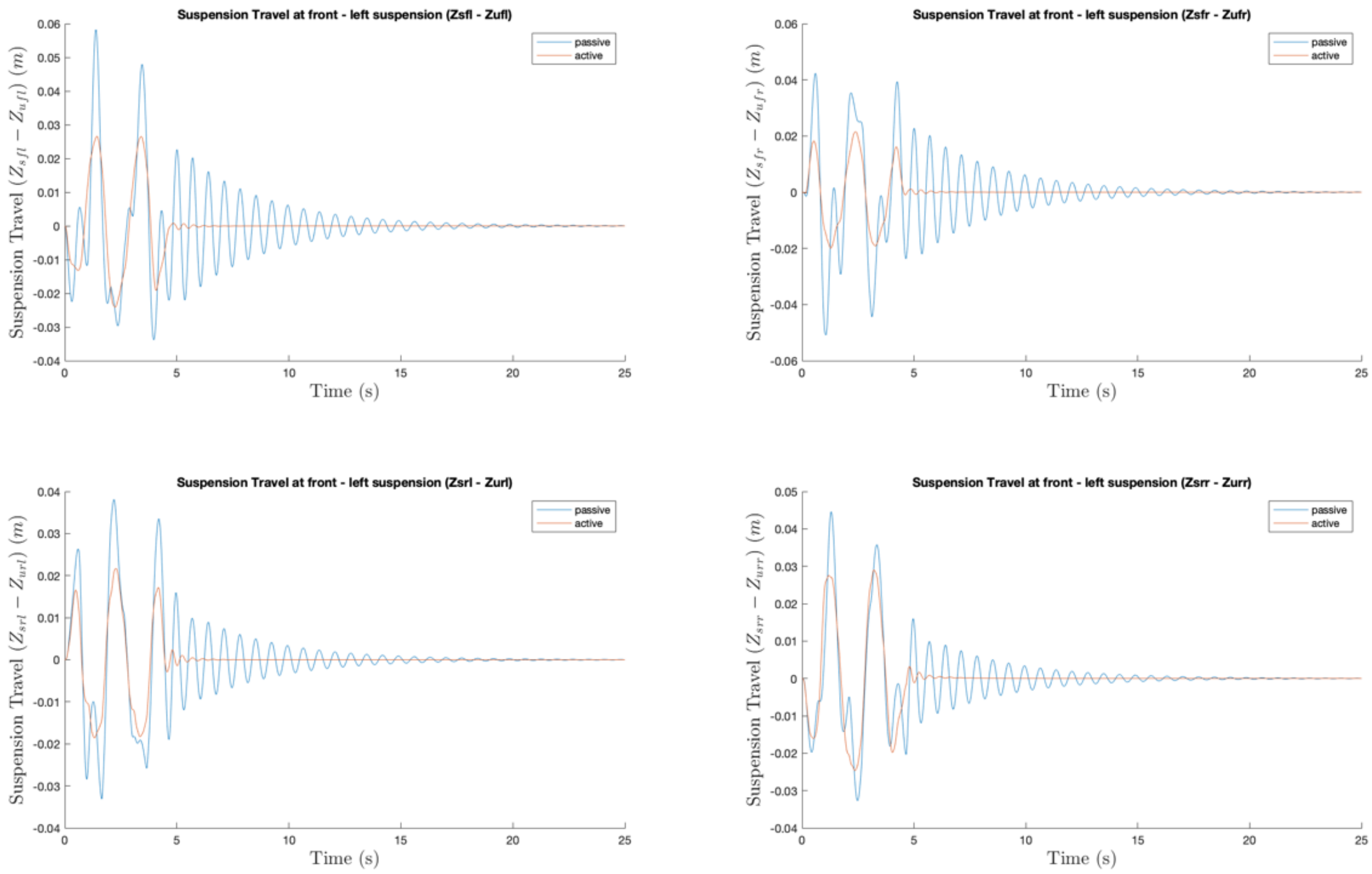

*Figure 8 (b): Plots of suspension travel vs time for the FC model (the four wheels are illustrated in the plots)*

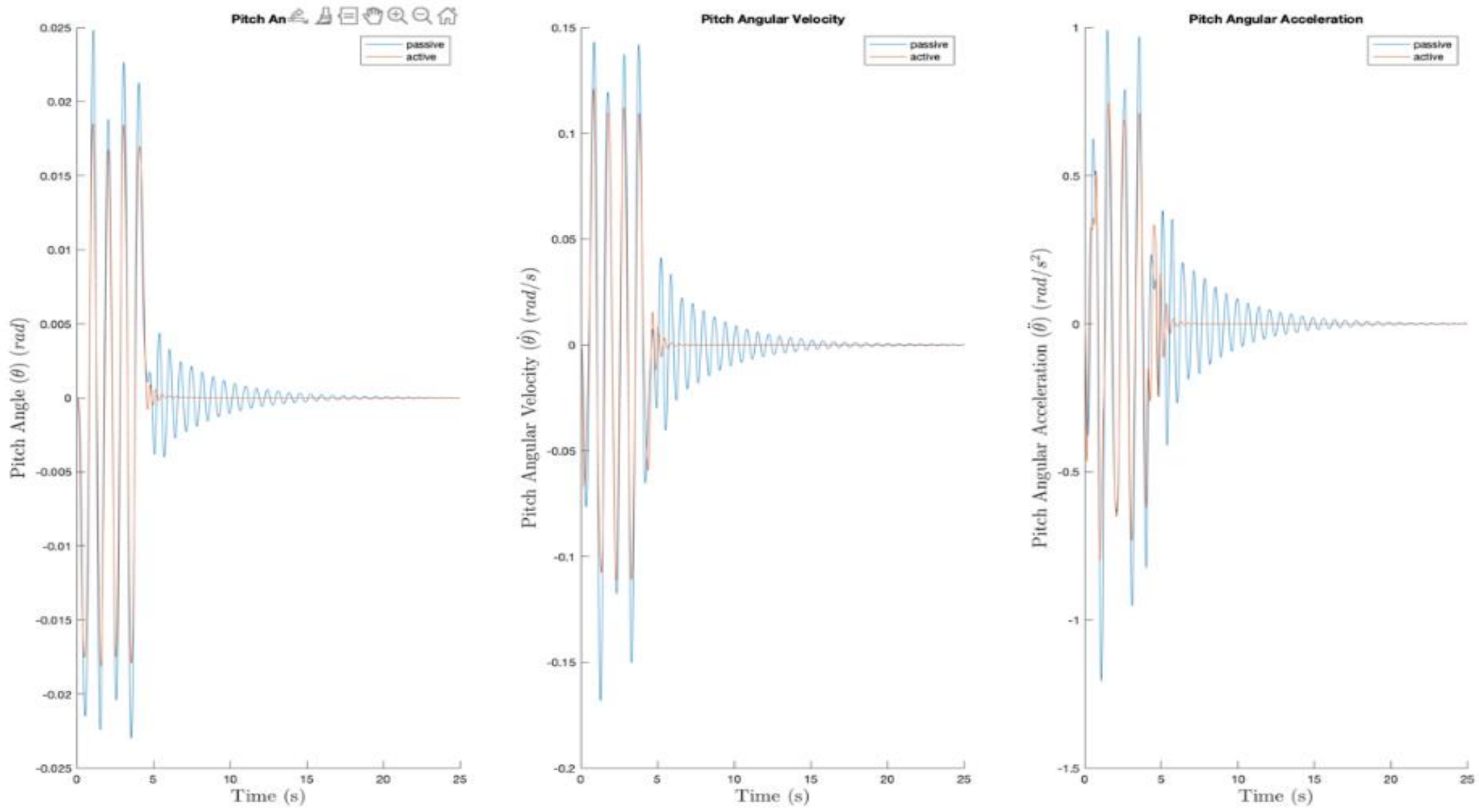

*Figure 8 (c): Plots of pitch angle vs time for the FC model*

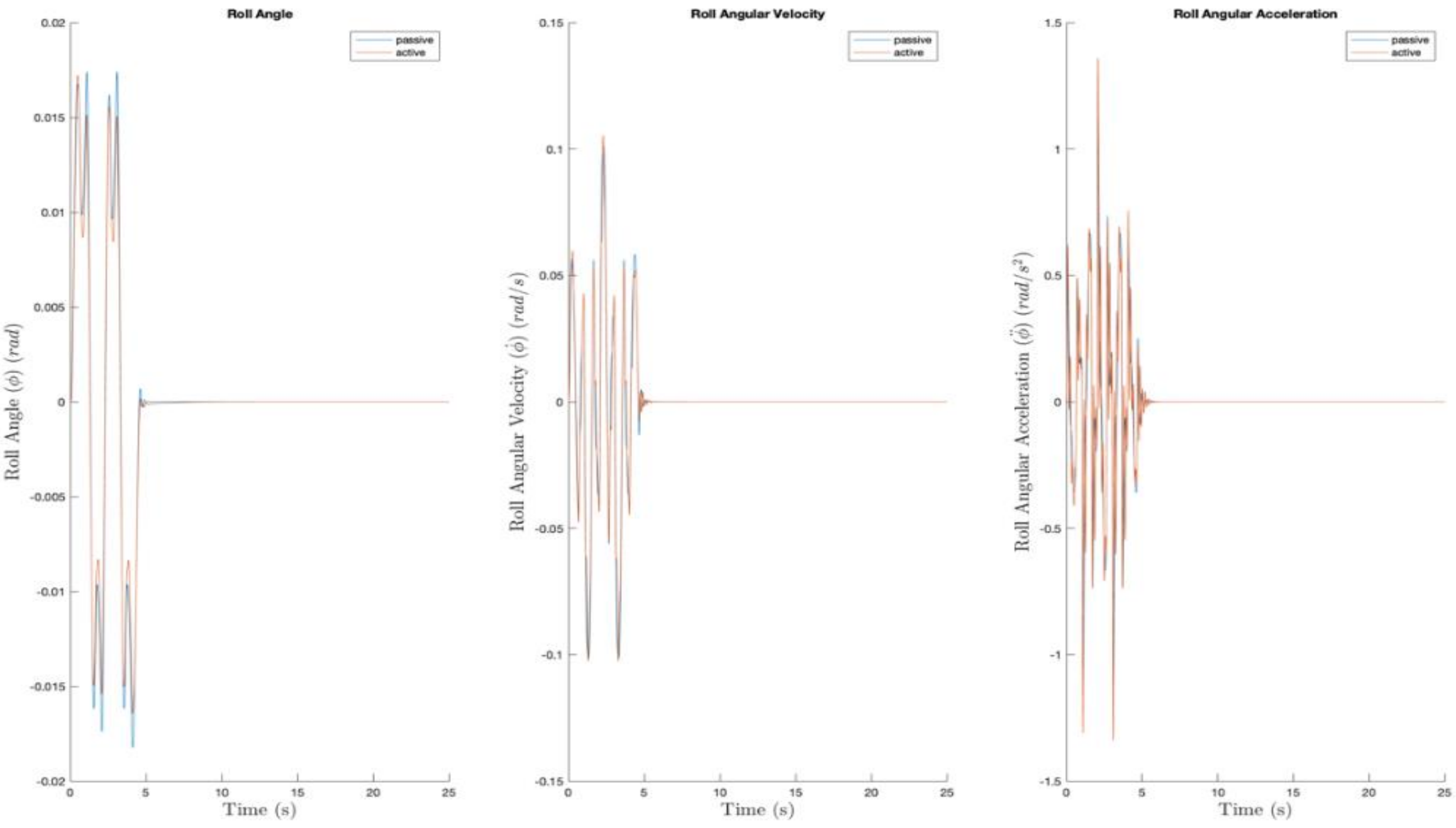

*Figure 8 (d) Plots of roll angle vs time for the FC model*

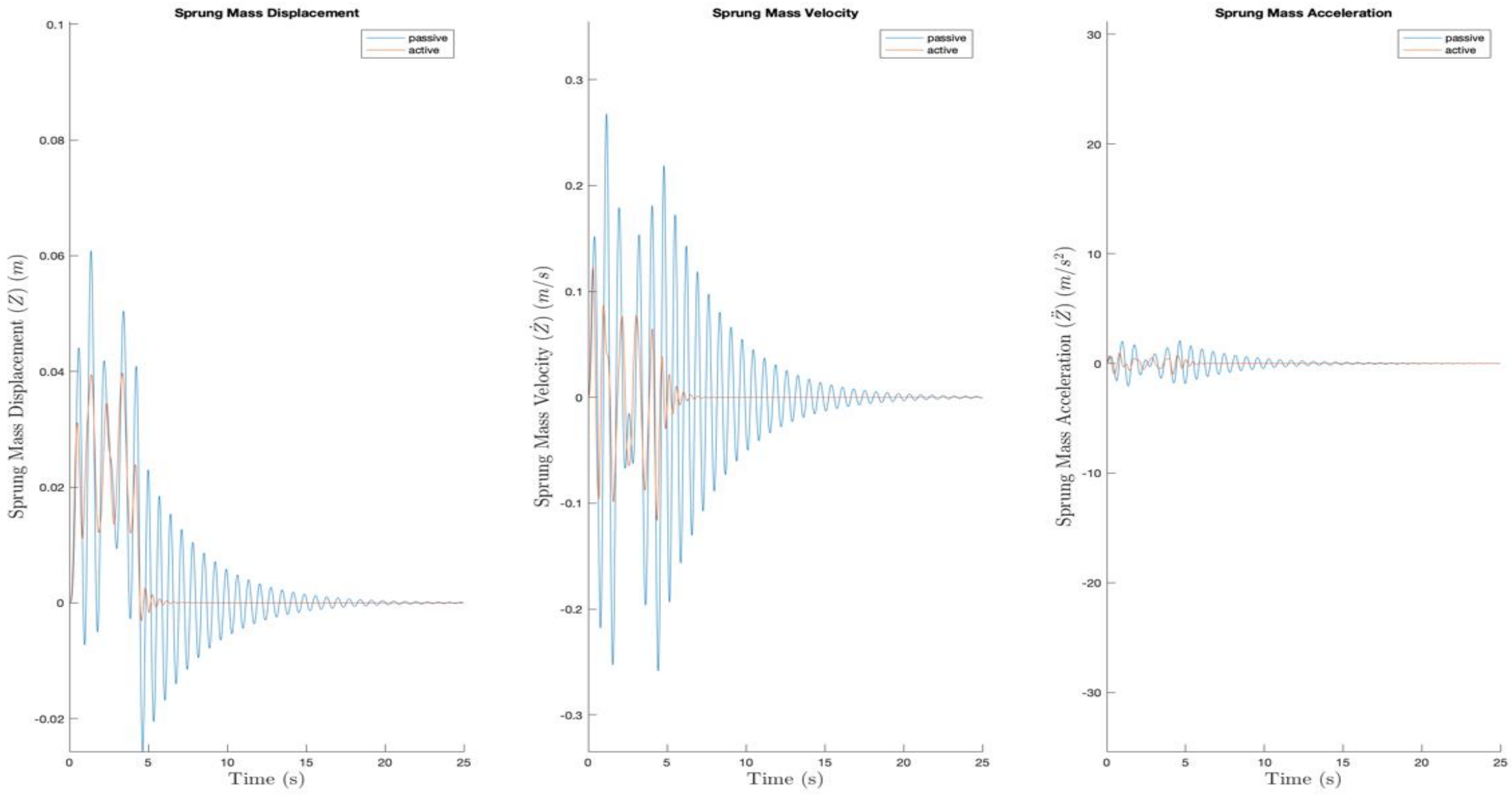

*Figure 8 (e): Plots of sprung mass displacement vs time for the FC model*

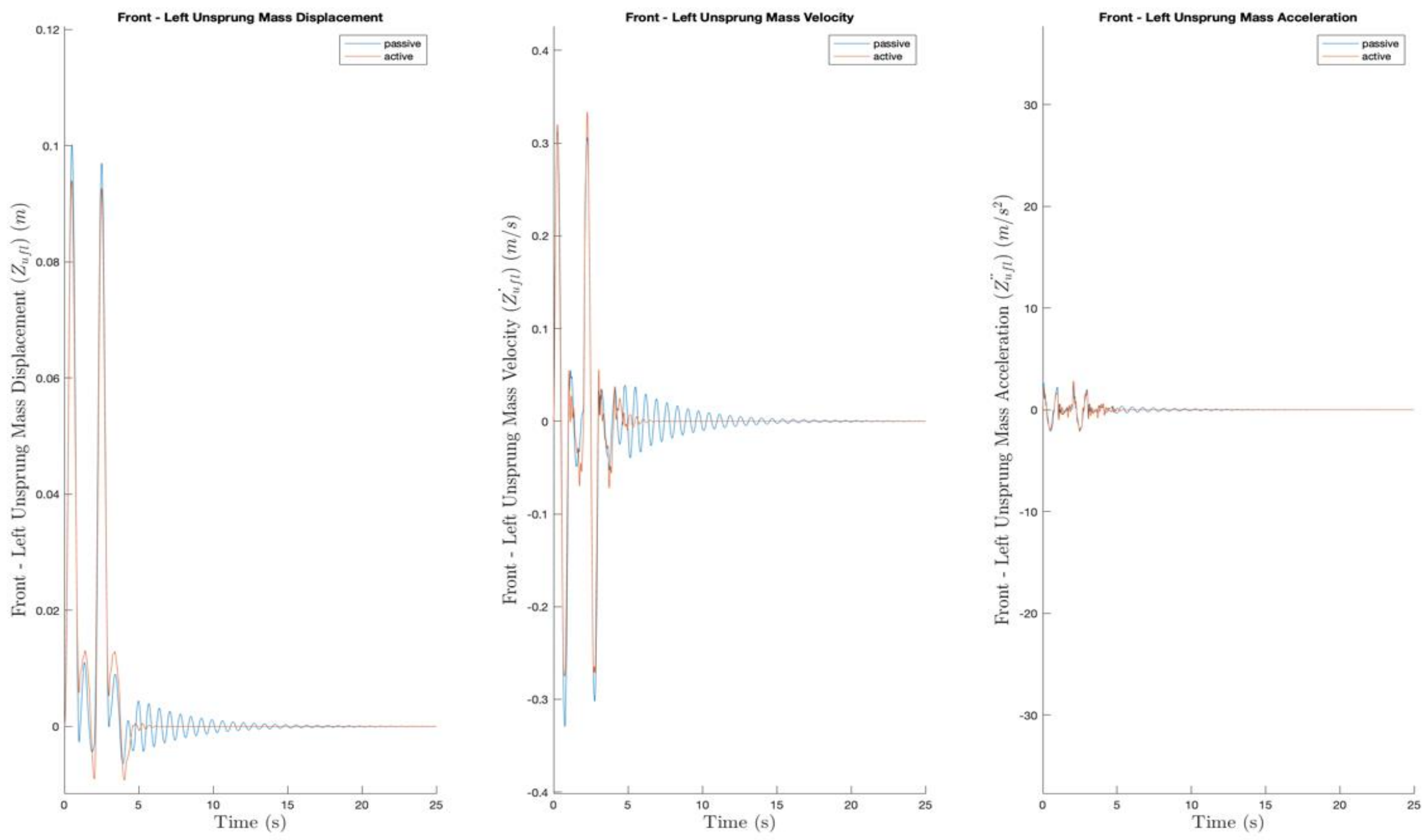

*Figure 8 (f1): Plots of unsprung mass displacement vs time for the FC model (Front-Left Wheel)*

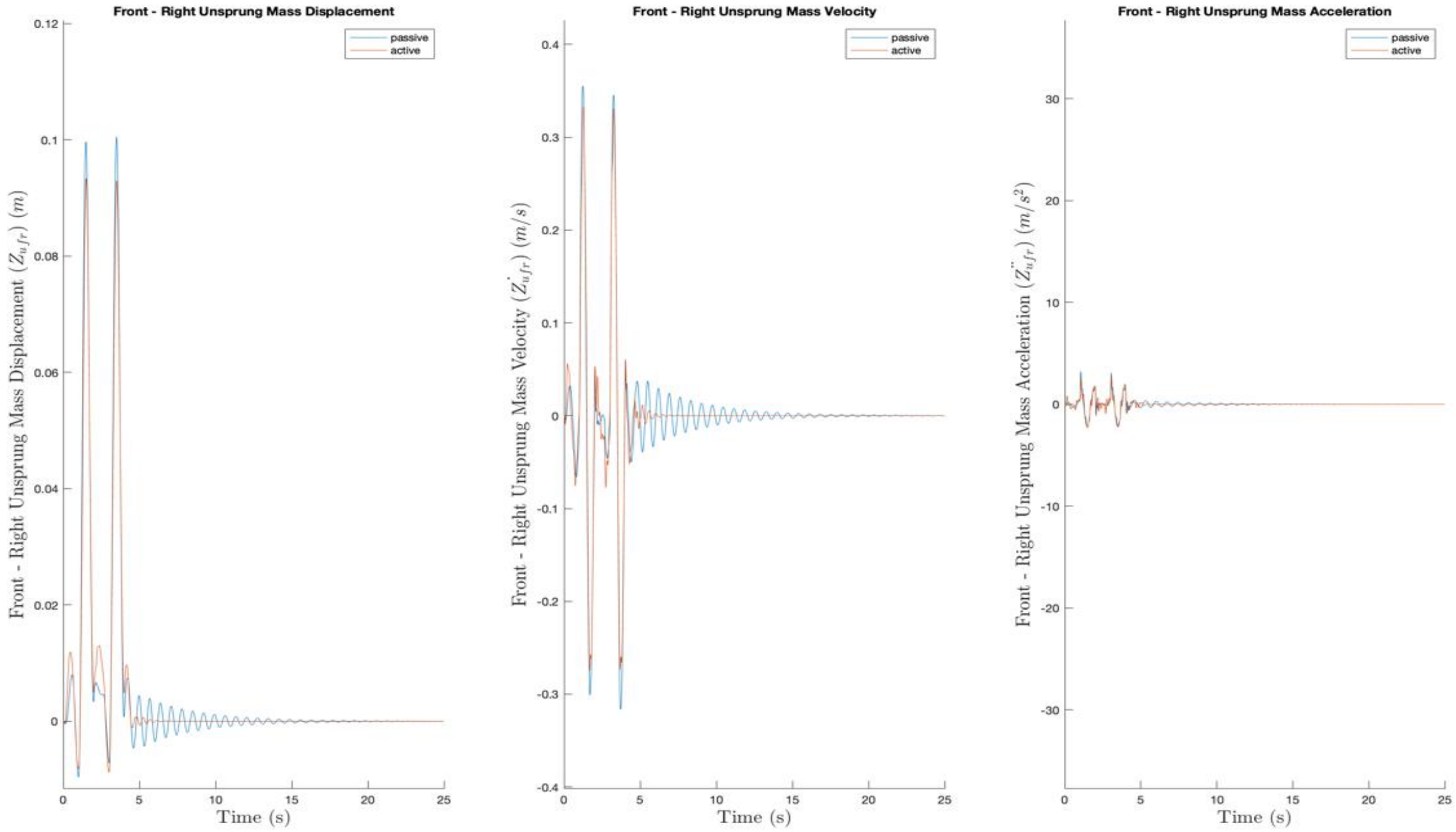

*Figure 8 (f2): Plots of unsprung mass displacement vs time for the FC model (Front-Right Wheel)*

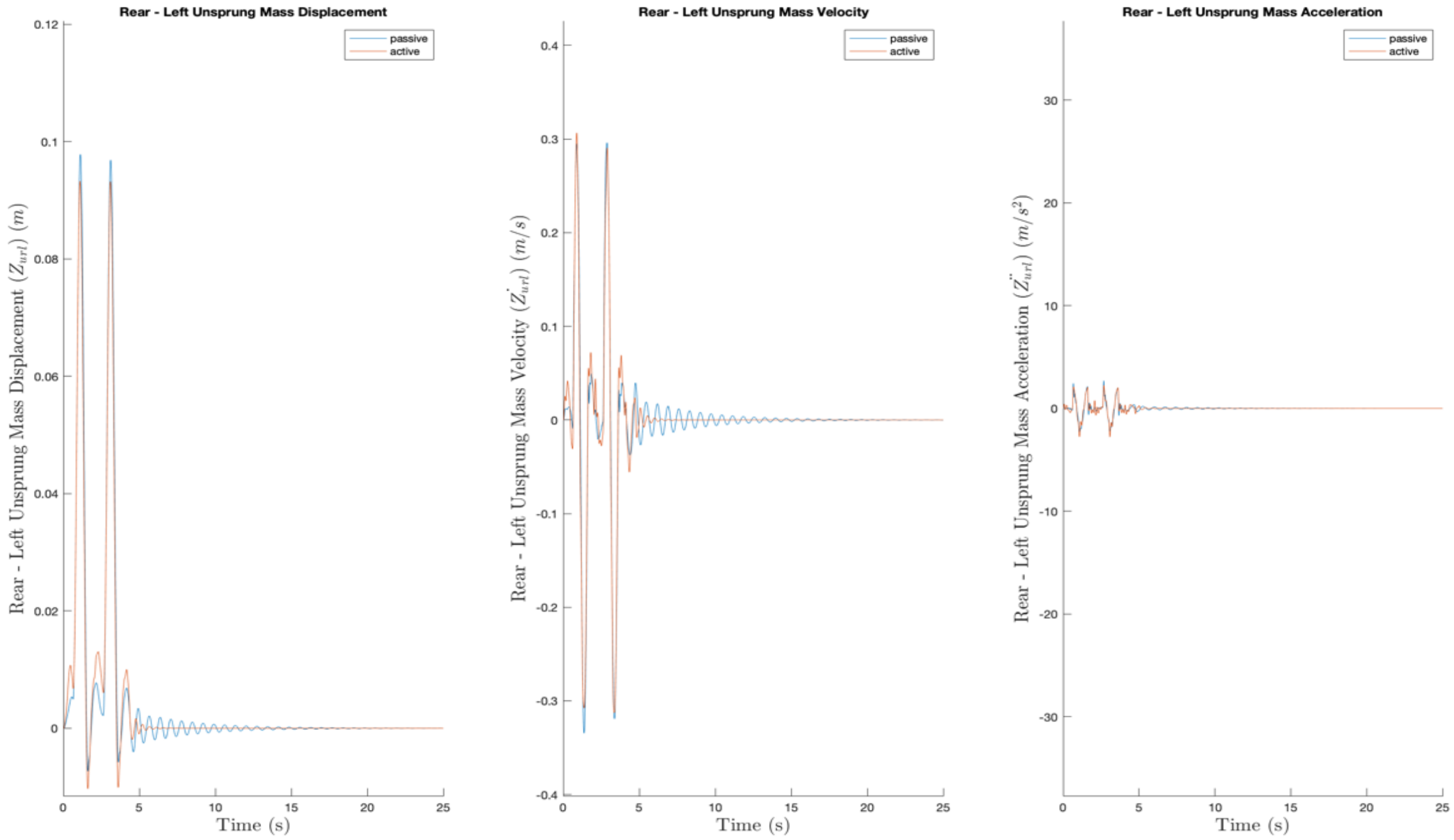

*Figure 8 (f3): Plots of unsprung mass displacement vs time for the FC model (Rear-Left Wheel)*

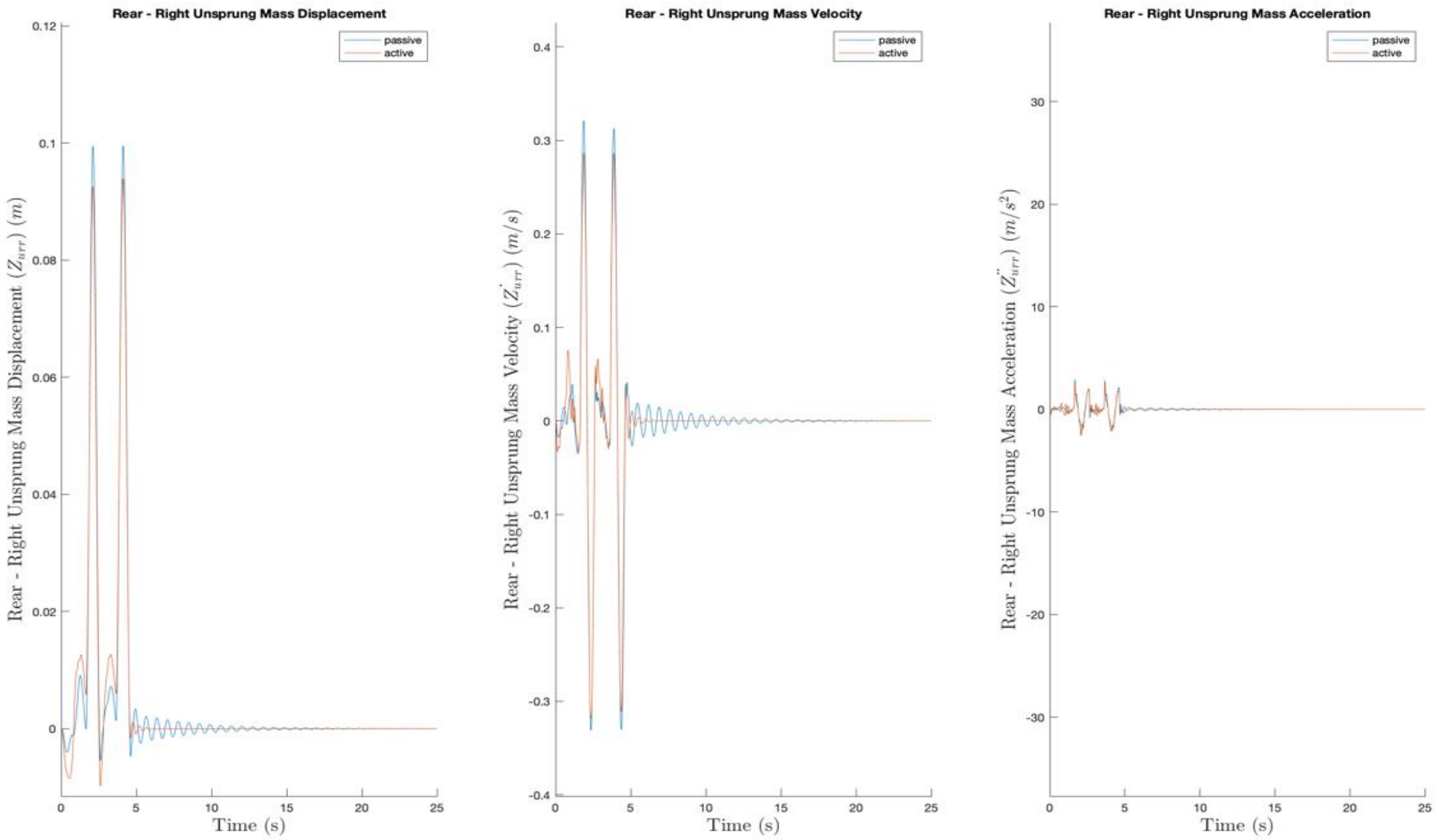

*Figure 8 (f4): Plots of unsprung mass displacement vs time for the FC model (Rear-Right Wheel)*

A comprehensive result of the full vehicle performance (after GA-LQR optimisation) in terms of suspension travel, sprung mass displacement, unsprung mass displacement, pitch angle and roll angle in the time domain is presented in Figure 8 (a)-(f). Figure 8 (a) shows the road profile used for the simulation. It was observed that the introduction of the active elements with GA-LQR optimised PID gains reduced the suspension travel (at the four wheels), the pitch angular acceleration and the sprung mass acceleration. However, no significant differences were observed between the roll angular acceleration and the unsprung mass accelerations (at the four wheels) of the PSS and the optimised active elements. A reduction from around 20 seconds to 6 seconds can be observed in the VST for the full car optimised ASS as compared to the PSS. Similar performance of the ASS over the PSS has been observed in the case of quarter car model results. A considerable reduction in the sprung mass displacement can be observed in Figure 8 (e). Although the unsprung mass acceleration at the four wheels did not reflect any significant change between the PSS and the ASS, the overall displacement of the unsprung mass of the ASS is lower than that of the PSS. This indicates an appreciable road holding. With regards to pitch angular acceleration furnished in Figure 8 (c), for the ASS, the acceleration amplitude range is lower and consequently returns to zero very fast at about 6 sec. This is an indication that the LQR objective function can be successfully integrated with GA for enhanced performance of an ASS equipped with a PIDC. Similar results have been documented by Shirahatti et al., [25].

### 5.4. Full Car Model with GA-CB Optimization for the PID Parameters

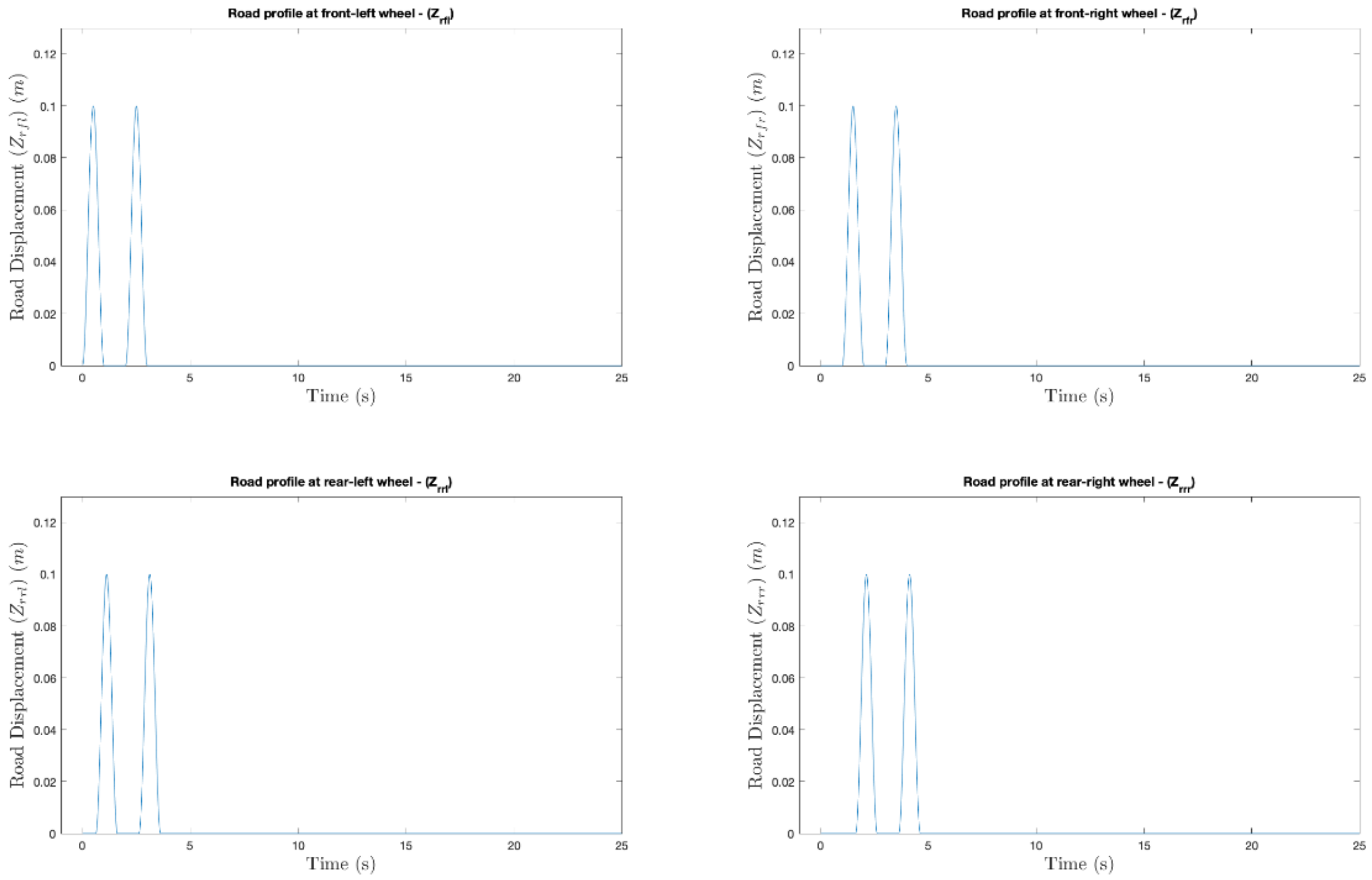

*Figure 9 (a): Plots of road displacement vs time for the FC model (the four wheels are illustrated in the plots)*

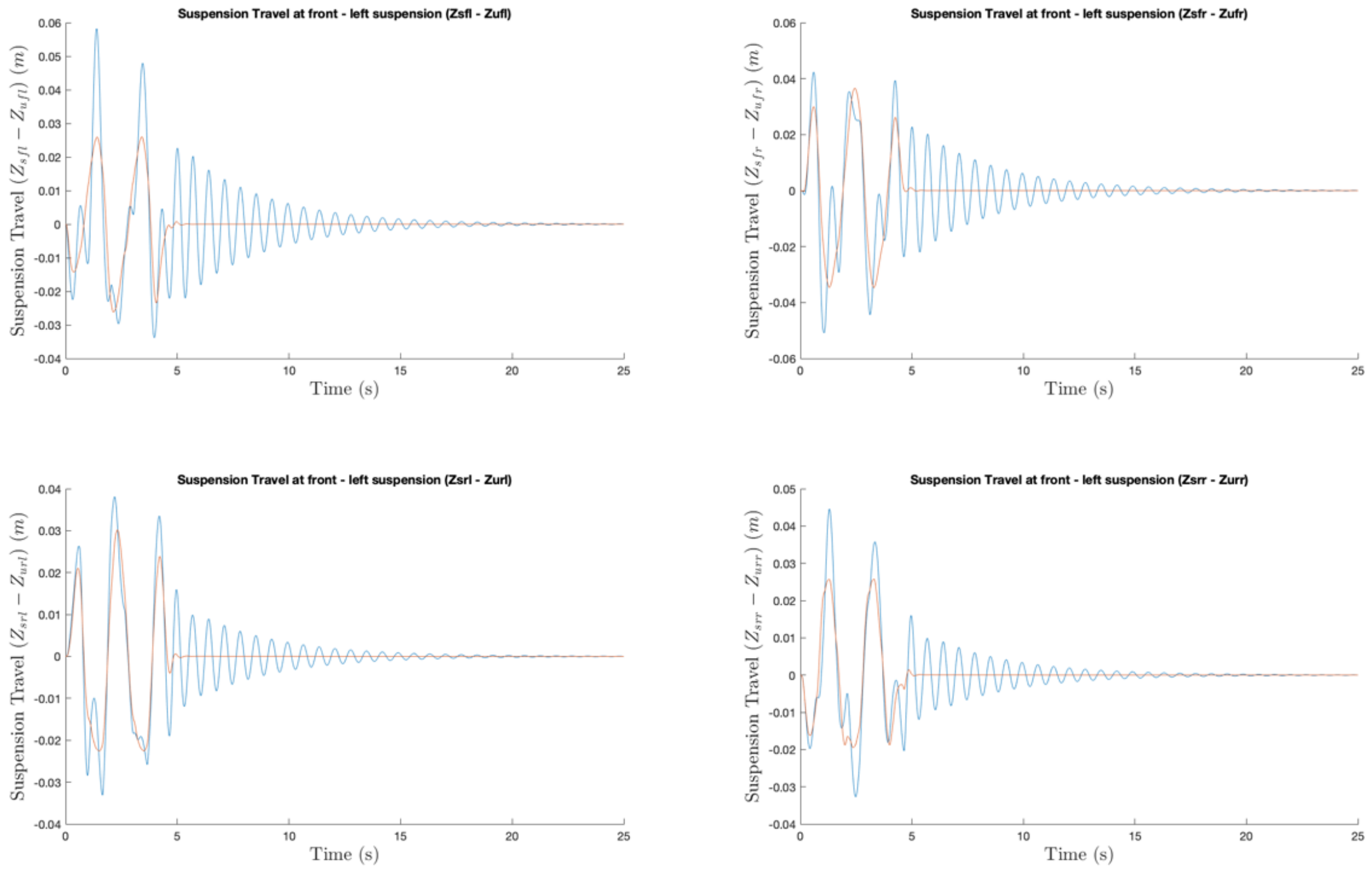

*Figure 9 (b): Plots of suspension travel vs time for the FC model (the four wheels are illustrated in the plots)*

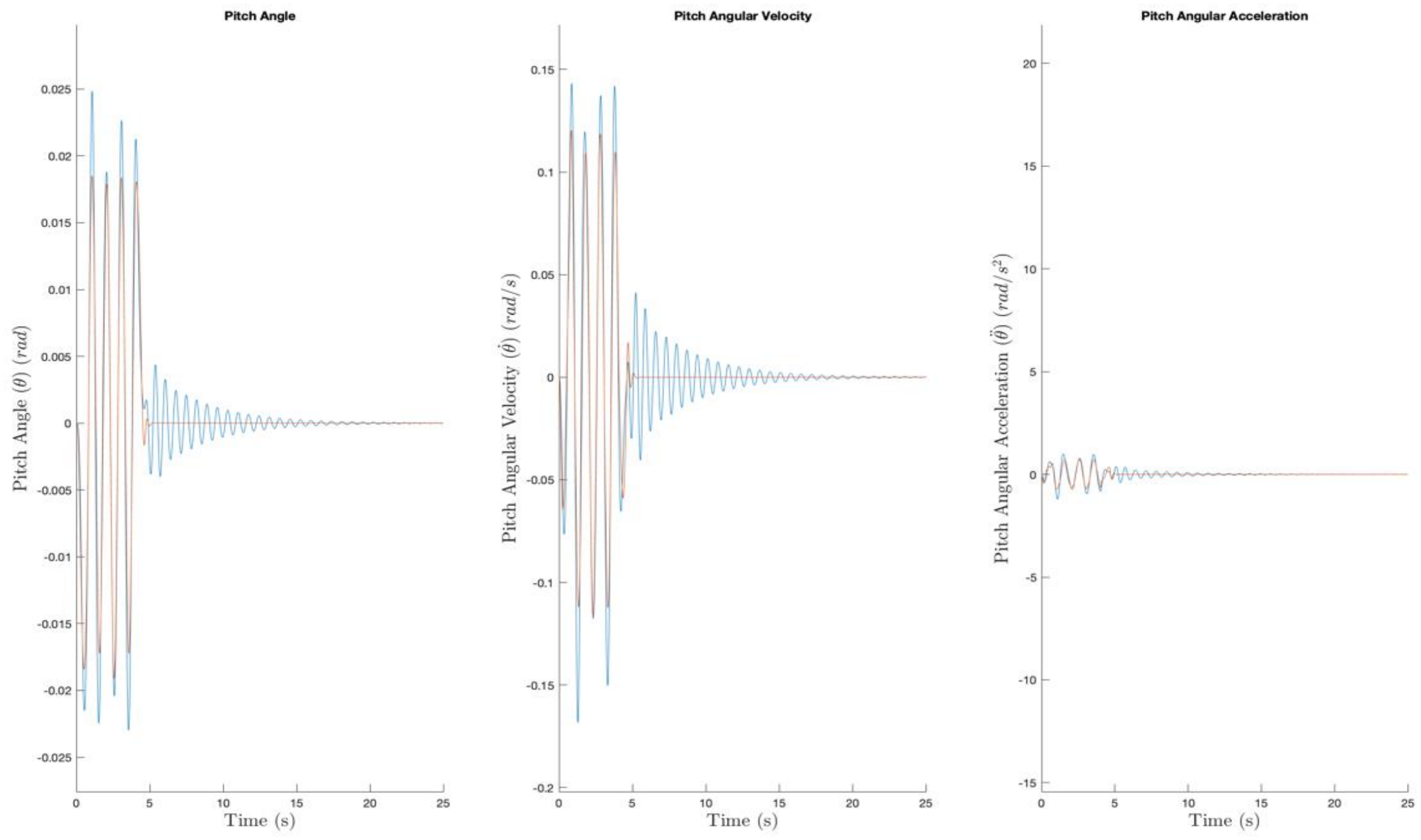

*Figure 9 (c): Plots of pitch angle vs time for the FC model*

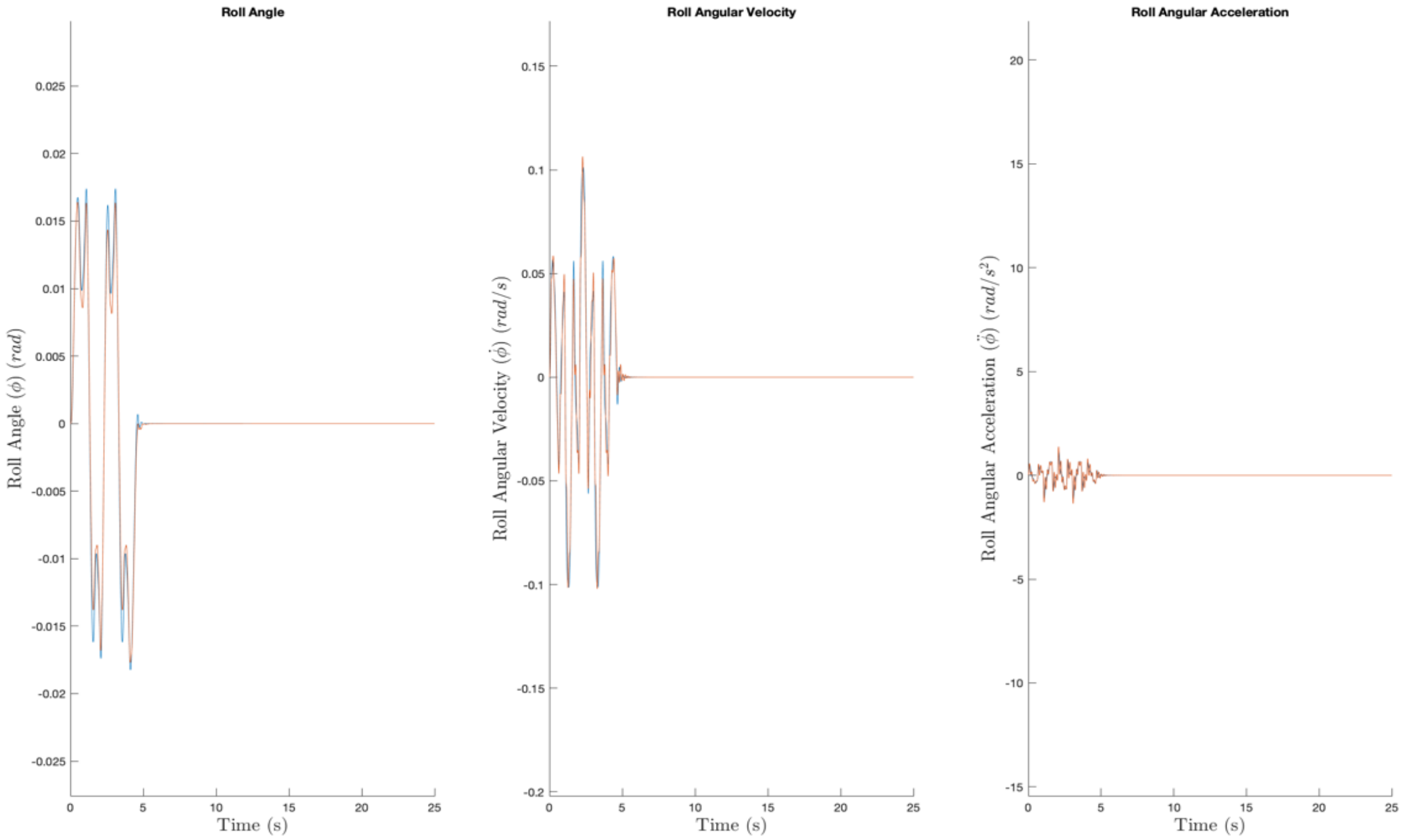

*Figure 9 (d): Plots of roll angle vs time for the FC model*

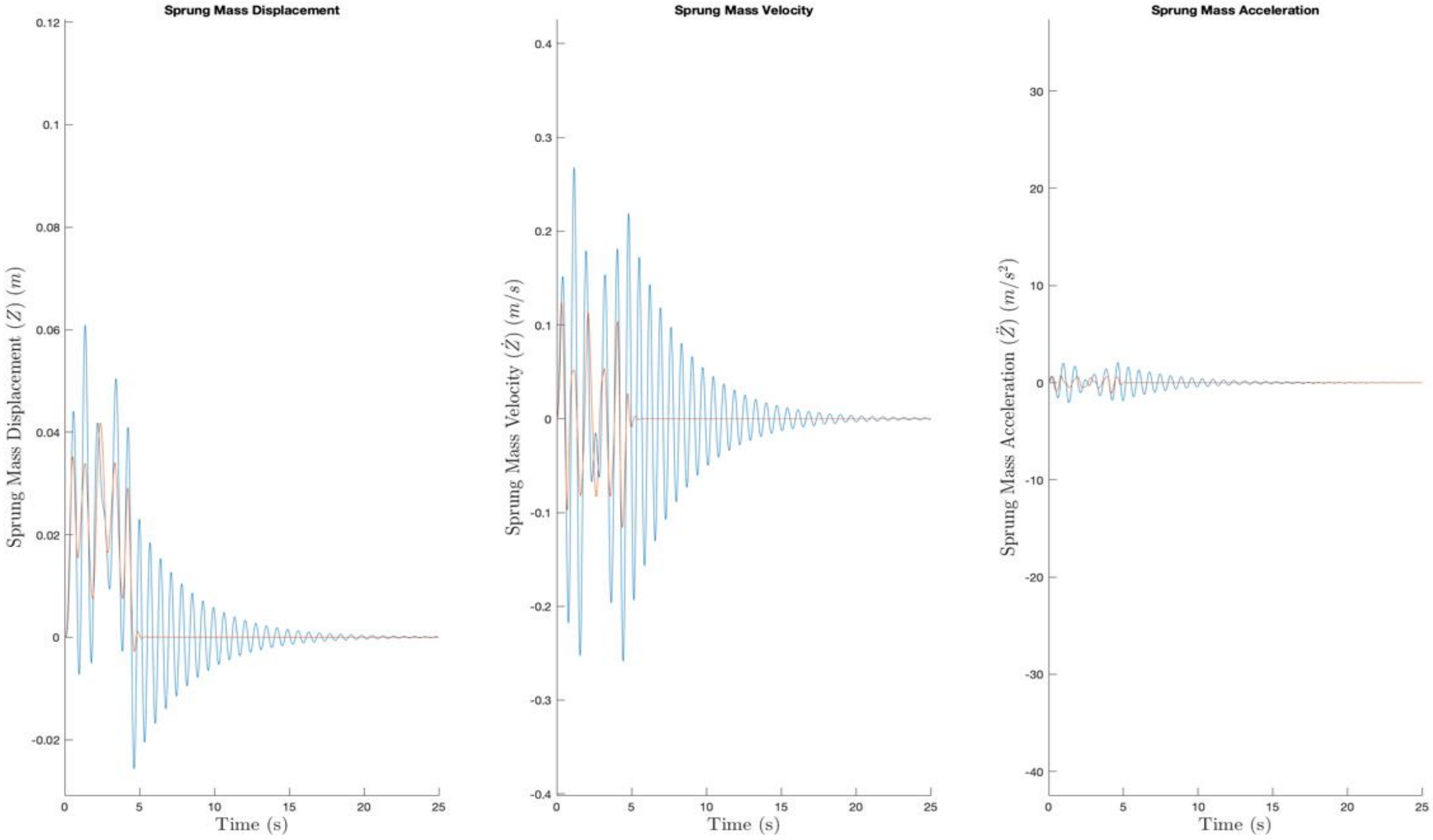

*Figure 9 (e): Plots of sprung mass displacement vs time for the FC model*

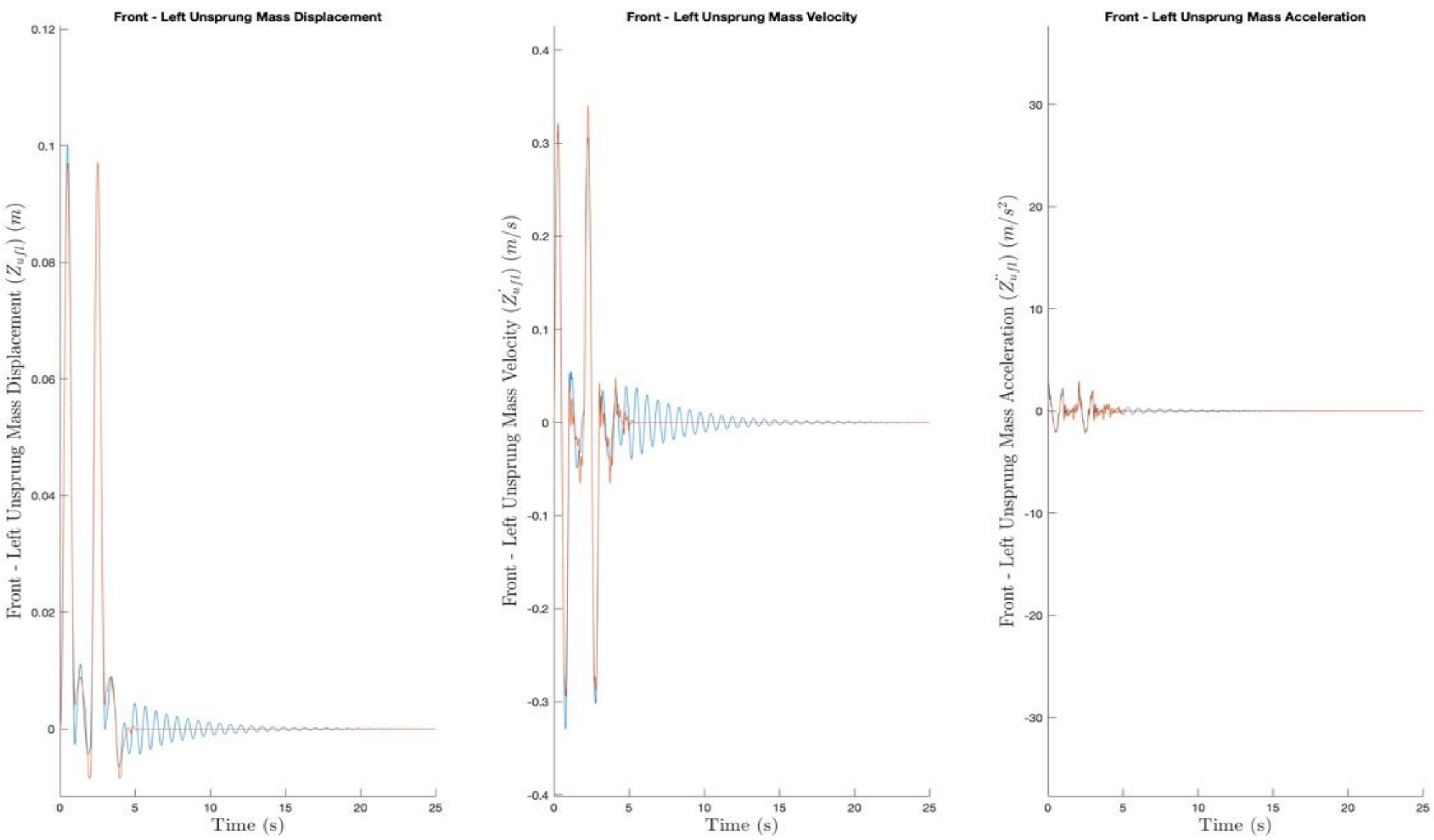

*Figure 9 (f1): Plots of unsprung mass displacement vs time for the FC model (Front-Left Wheel)*

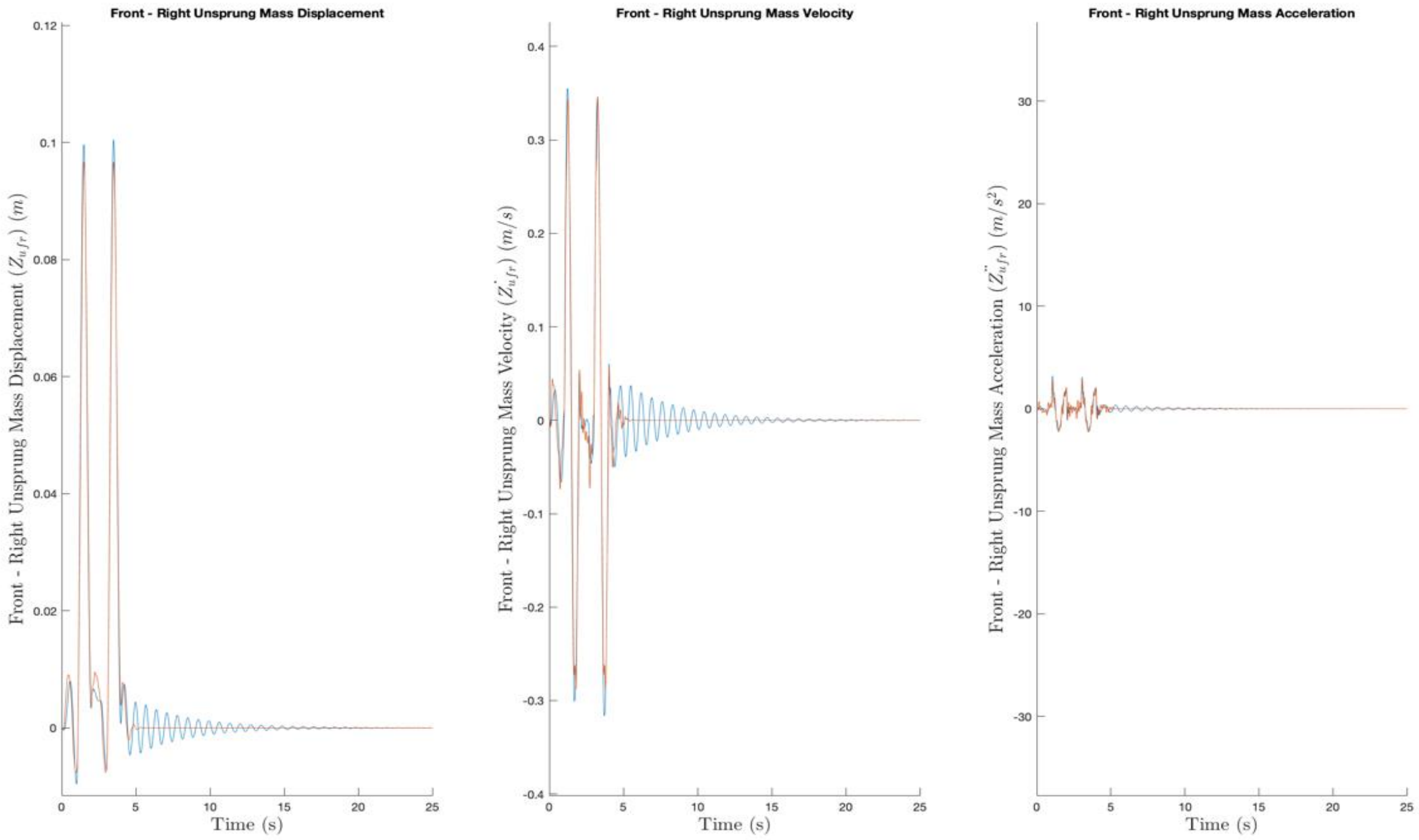

*Figure 9 (f2): Plots of unsprung mass displacement vs time for the FC model (Front-Right Wheel)*

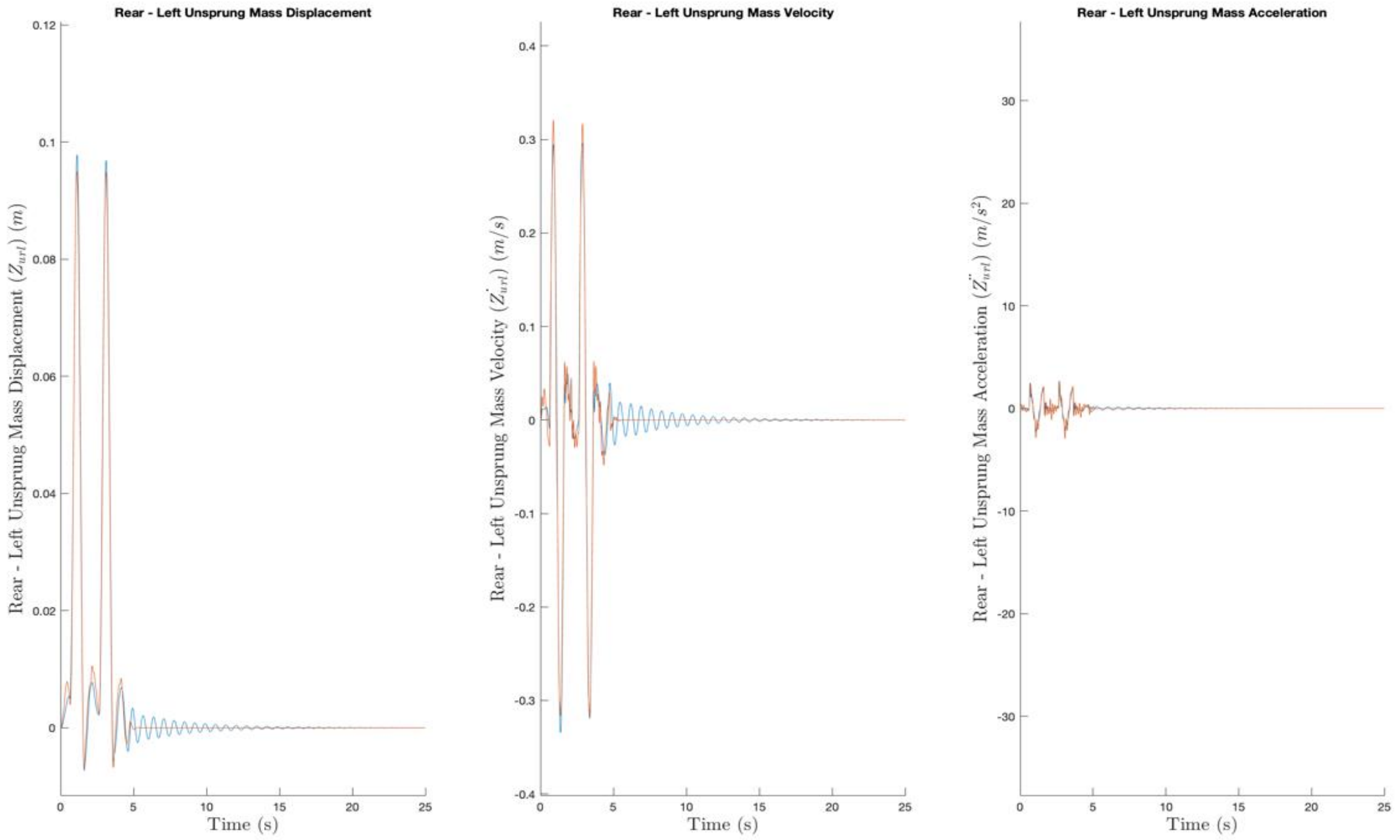

*Figure 9 (f3): Plots of unsprung mass displacement vs time for the FC model (Rear-Left Wheel)*

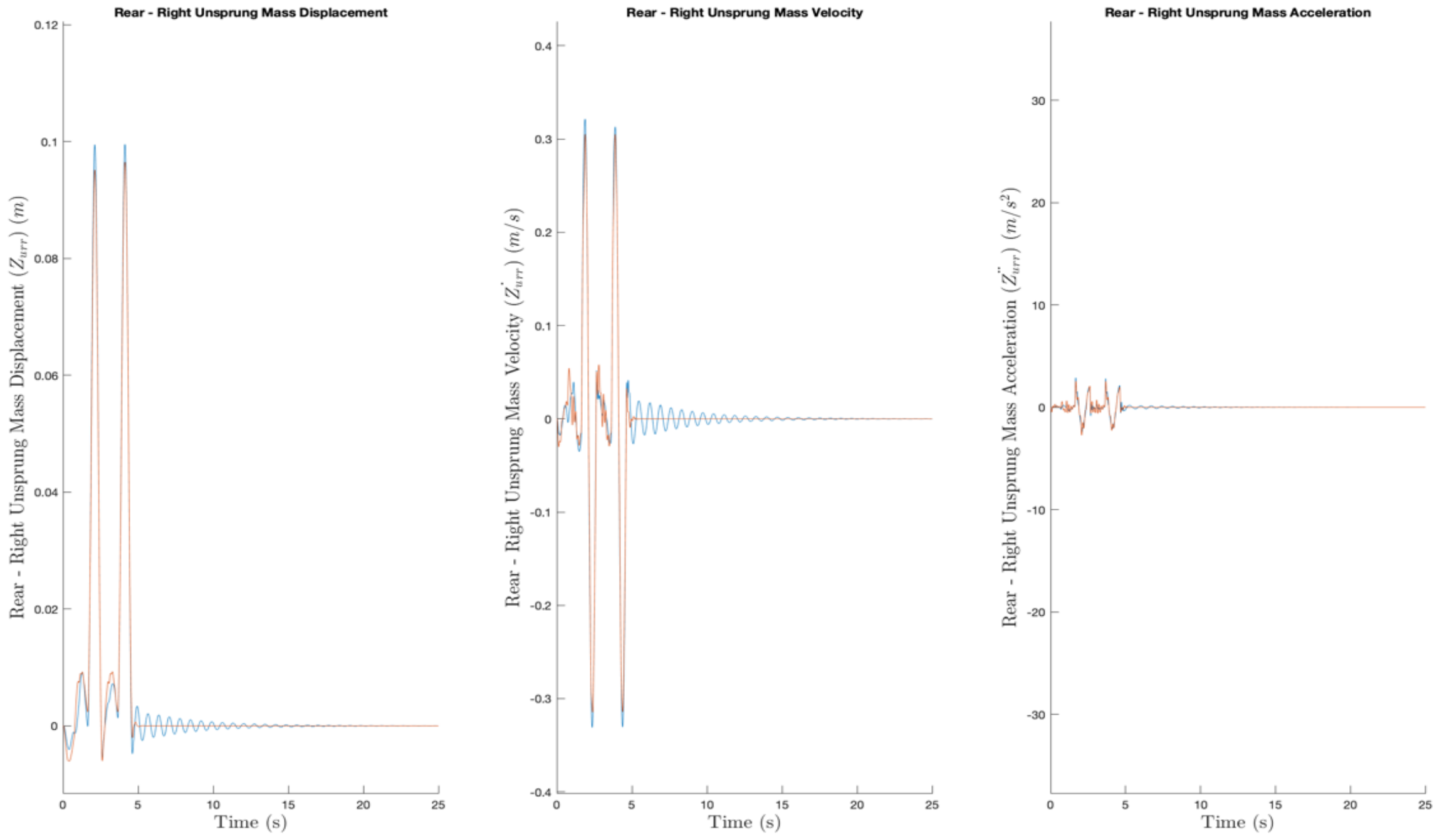

*Figure 9 (f4): Plots of unsprung mass displacement vs time for the FC model (Rear-Right Wheel)*

Figure 9 (a) shows the road profile used for the simulated results. Similarly, the GA-CB optimisation showed similar results (Figure 9) as was for observed for the GA-LQR optimisation in terms of suspension travel, sprung mass displacement, unsprung mass displacement, pitch angle and the roll angle in the time domain. For this case, it was also observed that the introduction of the active elements with GA-CB optimised PID parameters (PID gains inclusive) reduced the suspension travel (at the four wheels), the pitch angular acceleration, the pitch angle, the pitch angular velocity, and the sprung mass acceleration. However, no significant differences were observed between the roll angular acceleration and the unsprung mass accelerations (at the four wheels) of the PSS and the optimised active elements. . The VST reduced from 20 to 5 seconds for the fully optimised ASS case as compared to the PSS case. The vibrations die down as soon as the rear wheel passes the second bump. Also, the maximum displacement of the sprung mass has reduced by more than 30% for the fully optimised ASS. Although the unsprung mass acceleration at the four wheels did not reflect any significant change between the PSS and the ASS, the overall displacement of the unsprung mass of the ASS is lower than that of the PSS. This indicates an appreciable road holding compared to the PSS. With regards to pitch angular acceleration furnished in Figure 9 (c), for the ASS, the acceleration amplitude range is lower and consequently returns to zero very fast at about 6 sec. This is an indication that the CB objective function can be successfully integrated with GA for enhanced performance of an ASS equipped with a PID controller. Similar results have been documented by Shirahatii et al., [25]. Table 4 presents the PID gains for controller placed at wheels at front left, front right, rear left and rear right for the GA-LQR and GA-CB optimization approaches. Table 4 summaries the optimised suspension parameters, i.e., the suspension stiffness and the damping coefficient for the quarter and full car model.

*Table 4: Optimised PID gains for the controller at four wheels with GA-LQR and GA-CB for full car model*

| PID gains | GA-LQR | GA-CB |
|---|---|---|
| $Kp_{fl}$ | 54925 | 28001 |
| $Ki_{fl}$ | 14983 | 7261 |
| $Kd_{fl}$ | 4110 | 3587 |
| $Kp_{fr}$ | 77573 | 6793 |
| $Ki_{fr}$ | 28774 | 12257 |
| $Kd_{fr}$ | 6119 | 6802 |
| $Kp_{rl}$ | 75900 | 28162 |
| $Ki_{rl}$ | 35196 | 11848 |
| $Kd_{rl}$ | 1509 | 6694 |
| $Kp_{rr}$ | 44412 | 38241 |
| $Ki_{rr}$ | 61465 | 17675 |
| $Kd_{rr}$ | 1491 | 1673 |

*Table 5: Optimised suspensions parameters for half car and full car models*

| Parameter | Quarter Car | | Full Car | |
|---|---|---|---|---|
| Spring stiffness for front suspension (N/m) | $K_s$ | 73462 | $K_{sf}$ | 33786 |
| Spring stiffness for rear suspension (N/m) | - | - | $K_{sr}$ | 28037 |
| Damping coefficient for front suspension (N/m/s) | $C_s$ | 2578 | $C_{sf}$ | 1553 |
| Damping coefficient for rear suspension (N/m/s) | - | - | $C_{sr}$ | 1424 |

## Conclusions

The investigation starts with the understanding of the suspension system using the passive quarter car model. The response of this model shows the typical behaviour of the passive suspension system where the amplitude of displacements, velocities and accelerations for sprung and unsprung masses are high. Also, the system takes a long time to settle to the zero position after the initial stimulus in the form of a bump. From this understanding, the formulation including the active component to restrict the motion of the suspension in a controlled manner, was developed for the quarter car model. The active suspension system with an actuator controlled by the PID controller provides a better response than a passive system giving motivation the further investigation. Although the manually tuned PID controllers show promising results, their full potential can be harnessed to achieve a better ride comfort along with the conflicting road-handling criteria. This type of optimisation is only possible with the help of sophisticated algorithms like genetic algorithm.

The genetic algorithm uses the objective function defined for the LQR controller to obtain the optimised PID parameters. The manual aspect of the selection of penalty variables provides better control over the optimisation process; however, this makes it crucial to select these penalties carefully. The optimised quarter car active suspension model shows a significant improvement in ride quality.

The quarter car model provides insight into the behaviour of the suspension system. However, it is primitive in the sense that it cannot provide a detailed response of the existing suspension system of the vehicle, which includes four sets of suspension units. The full car model is the obvious next step to understand the response of suspension system including the bounce, pitch and roll. The full car passive suspension model was developed using the state-space formulation. The double bump road profile is used throughout this study to illustrate the road condition. A delayed bump on the right side of the vehicle compared to the left side induces the rolling and pitching motion to the vehicle providing valuable insight into the suspension behaviour. The response of this passive system defines the baseline response for the vehicle suspension system and the improvement achieved by the active suspension system can be judged based on this response.

Similar to the quarter car model, the genetic algorithm provides the optimisation solution to get PID parameters for achieving a better response. For the optimised system, the amplitudes of the displacements, velocities and accelerations are reduced considerably. One thing noted at this stage was that the number of PID gains are four times more than the quarter car case. Hence the manual selection of the penalty values becomes more difficult and calls for a better objective function and constraint handling strategy. This issue was addressed in the next stage.

The second branch of the optimisation was to obtain optimal spring stiffnesses and damping coefficients. The passive quarter and full car models are optimised to achieve this objective. These systems show significant improvement over the un-optimised passive suspension models. The objective function used for this purpose was different from the LQR cost function, which involved the minimisation of the sprung mass acceleration as the prime objective. The advantage of using this type of objective function is that the constrained problem is transformed into an unconstrained one. Also, unlike the LQR cost function, only one penalty parameter is to be selected, which reduces the human component from the optimisation process and the Genetic Algorithm can be exploited to the full extent.

The final stage of the study was to combine the understanding gained in the previous stages to optimise the spring stiffnesses, damping coefficients and PID gains simultaneously. The quarter car model was optimised with these objectives using the previously formulated objective function with the constraints. This model shows the best performance amongst the quarter car models pertaining to the availability of more control parameters to converge to the optimal solution.

In the continuation, the full car model undertakes the same objectives to obtain a fully optimised model. Also, the objective function utilised in the quarter car model provided the best performance leading to its use in the full car model. The sprung mass acceleration constraints provide control over the maximum as well as RMS acceleration, resulting in better rider comfort. On the other hand, road handling is taken care of by the unsprung mass displacement constraint. The tyre deflection constraints help in reducing the dynamic forces of the tyres, reducing the damage to the road. The suspension travel constraints help in reducing the possibility of suspension hitting the suspension stops during the operation. As there are frequency constraints embedded in the formulation, the comfortable ride frequency as well as the roll and pitch frequencies are assured. The added constraint on the jerk experienced by the occupants of the vehicle reduces the sudden changes in the acceleration. The constraints applied to the actuating forces as any actuator can provide a limited amount of force, and also it minimises the energy used for the actuation.

In short, from the results of the conducted in this study, it can be concluded that the active suspension system with optimised spring stiffnesses, damping coefficients and PID gains provides the best rider comfort and road handling balance and the genetic algorithm can be used to perform a successful optimisation.